\documentclass[structabstract]{aa}  
\usepackage{graphicx}
\usepackage{txfonts}
\usepackage{epsfig}
\usepackage{natbib}
\usepackage{lscape}
\usepackage{wasysym}
\usepackage{array}
\usepackage{amssymb}
\usepackage{subfigure}
\usepackage{longtable}
\usepackage{hyperref}
\usepackage{rotating}
\usepackage{color}
\usepackage{colortbl}
\usepackage{soul}
\usepackage[normalem]{ulem}
\usepackage{fancyheadings}
\bibpunct{(}{)}{;}{a}{}{,}

\begin{document}
   \title{Metallicity of M dwarfs}

   \subtitle{III. Planet-metallicity and planet-stellar mass correlations of the HARPS GTO M dwarf sample \thanks{Based on observations made with the HARPS instrument on the ESO 3.6-m telescope at La Silla Observatory under programme ID 072.C-0488(E)
   }}

\author{ V. Neves\inst{1,2,3} \and X. Bonfils\inst{2} \and
  N. C. Santos\inst{1,3} \and X. Delfosse\inst{2} \and
  T. Forveille\inst{2}  \and F. Allard\inst{4}  \and
  S. Udry\inst{5}}

\institute{
Centro de Astrof{\'\i}sica, Universidade do Porto, Rua das Estrelas,
4150-762 Porto, Portugal \\
email: {\tt vasco.neves@astro.ua.pt}
\and
UJF-Grenoble 1 / CNRS-INSU, Institut de Plan\' etologie et
d'Astrophysique de Grenoble (IPAG) UMR 5274, Grenoble, F-38041,
France.
\and
Departamento de F\'{\i}sica e Astronomia, 
Faculdade de Ci\^{e}ncias, Universidade do Porto, 
Rua do Campo Alegre, 4169-007 Porto, Portugal
\and
Centre de Recherche Astrophysique de Lyon, UMR 5574: CNRS,
Universit\'e de Lyon, \'Ecole Normale Sup\'erieure de Lyon, 46 All\'ee
d'Italie, F-69364 Lyon Cedex 07, France
\and
Observatoire de Gen\`eve, Universit\'e de Gen\`eve, 51 Chemin des
Maillettes, 1290 Sauverny, Switzerland
}

   \date{Received XXX; accepted XXX}

% \abstract{}{}{}{}{} 
% 5 {} token are mandatory
 
  \abstract
  % context heading (optional)
  % {} leave it empty if necessary  
   {}
  % aims heading (mandatory)
   { The aim of this work is the study of the planet-metallicity and the planet-stellar mass correlations for M dwarfs from the HARPS GTO M dwarf subsample.}
  % methods heading (mandatory)
   { We use a new method that takes advantage of the HARPS high-resolution spectra to increase the precision of metallicity, using previous photometric calibrations of [Fe/H] and effective temperature as starting values.}
  % results heading (mandatory)
   { In this work we use our new calibration (rms = 0.08 dex) to study the planet-metallicity relation of our sample. The well-known correlation for Giant planet FGKM hosts with metallicity is present. Regarding Neptunians and smaller hosts no correlation is found but there is a hint that an anti-correlation with [Fe/H] may exist. We combined our sample with the California Planet Survey late-K and M-type dwarf sample to increase our statistics but found no new trends. 
   
We fitted a power law to the frequency histogram of the Jovian hosts for our sample and for the combined sample, $f_{p} = C10^{\alpha[Fe/H]}$, using two different approaches: a direct bin fitting and a bayesian fitting procedure. We obtained a value for $C$  between 0.02 and 0.04 and for $\alpha$ between 1.26 and 2.94.
   
Regarding stellar mass, an hypothetical correlation with planets was discovered, but was found to be the result of a detection bias.}
  % conclusions heading (optional), leave it empty if necessary 
   {}

\keywords{stars: fundamental parameters -- 
stars: late type --
stars: low mass --
stars: atmospheres --
stars: planetary systems
}

   \maketitle
%
%________________________________________________________________

\section{Introduction}
\label{intro}
%\textcolor{red}{Still don't know if I'm going to write a full introduction or just a resume of the previous paper and 3/4 paragraphs.}

Stellar mass and metallicity are two important observables directly connected to the formation and evolution of planetary systems. These quantities play an important role in core-accretion models of formation and evolution of planets, as shown by numerous works studying the relationship of both quantities with planet formation \citep[e.g.][]{Ida-2005,Kornet-2006, Kennedy-2008a, Thommes-2008, Alibert-2011, Mordasini-2012}. 

%The planets that orbit M dwarfs form in a different environment than solar-type and higher mass stars, as 

The initial conditions of planet formation (e.g. disk mass, temperature and density profiles, gravity, gas-dissipation and migration timescales) all change with stellar mass \citep[e.g.][]{Ida-2005, Kornet-2006, Kennedy-2008a, Alibert-2011}.  Metallicity also plays a major role in the efficiency of the formation of giant planets for FGK dwarfs, as shown by both models \citep[e.g.][]{Ida-2004b, Mordasini-2009, Mordasini-2012} and observational data in the form of a planet-metallicity correlation \citep[e.g.][]{Gonzalez-1997,Santos-2004b,Fischer-2005, Sousa-2011b, Mayor-2011}, that  seems to partially vanish for Neptunian and smaller planet hosts \citep[]{Sousa-2008,Bouchy-2009, Ghezzi-2010,Sousa-2011b,Buchhave-2012}.

According to \citet{Thommes-2008} and \citet{Mordasini-2012}, a lower metallicity can be compensated by a higher disk mass to allow giant planet formation (and vice-versa -- the so called `compensation effect'). This result implies that M dwarfs, which are expected to have a lower disk mass \citep[e.g.][]{Vorobyov-2008, Alibert-2011} can form giant planets, but only if they have high metallicities, thus suggesting an even stronger giant planet-metallicity correlation compared to FGK dwarfs.

Disk instability models \citep[e.g.][]{Boss-1997}, on the other hand, do not predict, in general, the dependence of the planet formation on metallicity \citep{Boss-2002} and they also don't seem to depend strongly on stellar mass, at least in the case of M dwarfs \citep{Boss-2006a}. %\textcolor{red}{Should we detail the feh dependent models? All of them Mayer(2007), Cai 2007, etc -- see Sarah Seager's book. fail to fragment. (!)}
%\textbf {or predict a slight dependence on [Fe/H] \citep[e.g.][]{} -TBD }. 
Contrary to the core-accretion paradigm \citep{Pollack-1996}, the formation of planets does not originate from the collisional accretion of planetesimals, but from the collapse of an unstable part of the protoplanetary disk, forming in a timescale of thousands of years when compared to a timescale of Myrs for core-accretion models. Observational evidence, however, has shown that there is a dependence between planet occurrence and both stellar mass and metallicity over a wide range of dwarf types \citep[AFGKM -- e.g.][]{Laws-2003,Bonfils-2007, Lovis-2007, Johnson-2007, Johnson-2010} , thus favoring the core-accretion scenario as the primary mechanism of planet formation, at least for closer-in planets. 

In this context, the `pollution' scenario \citep[e.g.][]{Gonzalez-1997, Murray-2002}, defends that the observed enhanced metallicity is only at the surface of the photosphere, and that the formation of planets occurs at all metallicities, thus supporting disk instability models. Observationally, this would translate, for M dwarfs into a non-detection of the planet-metallicity correlation, as M dwarfs have very deep convective layers and are expected to be fully convective at masses below 0.4 M$_{\odot}$. %However, as we have seen, most studies hint or demonstrate the presence of the metallicity correlation for giant planets, thus signaling a primordial origin of the presence of metals.

Recent observational works for M dwarfs are in line with a planet-metallicity correlation \citep[e.g.][]{Bonfils-2007,Johnson-2009, Schlaufman-2010, Rojas-Ayala-2012, Terrien-2012}. However, more detections of planets around M dwarfs and a more precise metallicity determination are needed to achieve higher confidence levels that remain low, below the $\sim$3~$\sigma$ level \citep{Bonfils-2007,Schlaufman-2010}. In this context it is important to use a volume-limited sample of stars, as several planet-hunting programs targeting FGK dwarfs have metallicity-biased samples \citep[e.g.][]{Baranne-1996,Fischer-2005b,Melo-2007}.

In the course of this paper we implement a new method to derive the metallicities of a volume-limited sample of 102 M dwarfs from the HARPS GTO programme. This method uses the high-resolution spectra of HARPS to achieve a [Fe/H] precision of 0.08 dex and is described in the Appendix. Then, we
%to refine and enhance the precision of metallicities based on the photometric calibration of \citet{Neves-2012}, itself a refinement of the work of\citet{Schlaufman-2010}. Then, we use this determinations, as well as stellar mass determinations calculated using the mass-luminosity relation of \citet{Delfosse-2000} 
search for correlations between the frequency of planets with stellar mass and metallicity. In Sect. \ref{sample}, we describe our M dwarf sample and observations using the HARPS spectrograph. Then, in Sect. \ref{relation}, we investigate the stellar mass/metallicity correlations with the frequency of planets. %Afterwards, we calculate the detection limits of the sample, to check for biases, and to test the stellar mass-planet relation.
%the planet detections accumulate in the higher end of stellar mass and lower end of V magnitude. 
Finally, we discuss our results in Sect. \ref{discussion}.

%The fact that ho 

%Efficiency of giant planets should increase with stellar mass (Laughlin 2004, Ida 2005b Kennedy 2007)

%No clear metallicity effect is found in the Neptunian mass domain and at lower masses even an anti-correlation might exist. (Mordassini 2012)

%Disk gas masses and giant planet masses are correlated (Mordassini 2012). 

%BUT Microlensing surveys show that 

%Mdisk is proportional to the Mdisk^alphad

%Alibert 2011 states that the number of embryos that eventually evolve and form planets are a growing function of the stellar mass. 

%The minimum metallicity required to form a massive planet is correspondingly lower for massive stars than lower mass stars. 
%
%However, \citet{Alibert-2011} states that the metallicity efffect is only weakly dependent on stellar mass. 

%--> Talk about the paucity of jovians around M dwarfs

%--> Should we talk about Disk instability?

%__________________________________________________________________

\section{Sample and Observations}
\label{sample}

Our sample of 102 M dwarfs is described in detail in Sect. 2 of \citet{Bonfils-2011}. It is a volume limited (11 pc) sample, containing stars with a declination $\delta< +20^{\circ}$, with V magnitudes brighter than 14  mag, and including only stars with a projected rotational velocity $vsini\le 6.5$ km/s. All known spectroscopic binaries and visual pairs with separation lower than 5 $arcsec$, as well as previously unknown fast rotators were removed \textit{a priori} or \textit{a posteriori} from the original sample. %All stars were surveyed for planets with the HARPS spectrograph \citep{Mayor-2003b}.

The observations were gathered using the HARPS instrument \citep{Mayor-2003b,Pepe-2004}, installed at the ESO 3.6-m telescope at the La Silla observatory in Chile. It is a high resolution (R$\sim$115000) spectrograph in the visible, covering a region between 380 and 690 nm. During the time of the GTO program, from 11th February 2003 to the 1st of April 2009, a total of 1965 spectra were recorded. The aim of the HARPS M dwarf programme is to achieve a $\sim 1$ m/s RV precision per exposure for the brightest targets. The chosen recording mode during this period was single fiber mode, that relies only on a single calibration but gives enough precision to reach the aim of the programme. Using single fiber mode has the advantage of obtaining non-contaminated spectra that can be used to perform studies other than measuring the star's RV, such as measuring activity diagnostics, using Ca II H and K lines, and calculating stellar parameters and abundances. A more detailed description of the observations is given in Sect. 3 of \citet{Bonfils-2011}.

From the 102 M dwarf stars, a total of 15 planets are currently detected, in 8 systems, from which 3 have more than one planet. Table \ref{planets} shows the planet hosts, planets, and planetary mass and period taken from \citet{Bonfils-2011}, except in the case  of Gl 876e \citep{Rivera-2010}. We refer to Table 1 of \citet{Bonfils-2011} for the full planet parameters and respective references.

%\addtocounter{table}{1}

\begin{table}[h]
\centering
\caption{Planet host stars in the sample, along with the planetary mass and period. We refer to \citet{Bonfils-2011} for the full references.}
\label{planets}
\begin{center}
%\resizebox{9cm}{!}{
\begin{tabular}{l l r r r }

\hline
\hline
star & planet & \multicolumn{2}{c}{$m\sin{i}^{\dag}$} & Period \\
     &        &  [M$_{\oplus}$] & [M$_{j}$]     & [days] \\
\hline
%Gl 176 & b &  8.4 & 0.026 &  8.7(8) \\
%Gl 433 & b &  6.4 & 0.0202 &  7.36(5) \\
%%Gl 433$^{1}$ & c & 44.6 & 0.14 & 3(693) \\
%Gl 581 & b & 15.7 & 0.0492 &  5.368(7)\\
%Gl 581 & c &  5.4 & 0.017 & 12.9(3) \\
%Gl 581 & d &  7.1 & 0.022 & 66.(8) \\
%Gl 581 & e &  1.9 & 0.0060 &  3.1(5) \\
%Gl 667C & b &  6.0 & 0.019 &  7.20(3) \\
%Gl 667C & c &  3.9 & 0.012 & 28.1(5) \\
%Gl 674 & b & 11 & 0.034 &  4.6(9) \\
%Gl 832 & b & 200 & 0.64 & 3(416) \\
%Gl 849 & b & 310 & 0.99 & 18(52) \\
%Gl 876 & b & 839 & 2.64 & 61.0(7) \\
%Gl 876 & c & 264 & 0.83 & 30.2(6) \\
%Gl 876 & d &  6.3 & 0.020 &  1.9378(5) \\
%Gl 876$^{2}$ & e & 14.6 & 0.046 & 124.(26) \\
Gl 176 & b &  8.4 & 0.026 &  8.78 \\
Gl 433 & b &  6.4 & 0.0202 &  7.365 \\
%Gl 433$^{1}$ & c & 44.6 & 0.14 & 3(693) \\
Gl 581 & b & 15.7 & 0.0492 &  5.3687\\
Gl 581 & c &  5.4 & 0.017 & 12.93 \\
Gl 581 & d &  7.1 & 0.022 & 66.8 \\
Gl 581 & e &  1.9 & 0.0060 &  3.15 \\
Gl 667C & b &  6.0 & 0.019 &  7.203 \\
Gl 667C & c &  3.9 & 0.012 & 28.15 \\
Gl 674 & b & 11 & 0.034 &  4.69 \\
Gl 832 & b & 200 & 0.64 & 3416 \\
Gl 849 & b & 310 & 0.99 & 1852 \\
Gl 876 & b & 839 & 2.64 & 61.07 \\
Gl 876 & c & 264 & 0.83 & 30.26 \\
Gl 876 & d &  6.3 & 0.020 &  1.93785 \\
Gl 876$^{2}$ & e & 14.6 & 0.046 & 124.26 \\

\hline
\hline
\end{tabular}
%}
\end{center}
\raggedright
$\dag$ The true mass (m$_{p}$) is reported for Gl876b,c \citep{Correia-2010}. \\
%$^{1}$ \citet{Delfosse-2012}\\
$^{2}$ \citet{Rivera-2010} 
\end{table}

%\textcolor{red}{should we put here the Table 1 of Bonfils(2011) but only including our sample? Should we put more information regarding the sample(observations? YES TBD.}

The stellar masses were calculated using the empirical mass-luminosity relationship of \citet{Delfosse-2000}, using stellar parallaxes, taken mostly from the HIPPARCOS catalogue \citep{van_Leeuwen-2007} , but also from \citet{van_Altena-1995, Jahreiss-1997, Hawley-1997, Henry-2006}. The V band magnitudes were taken from the Sinbad database\footnote{http://simbad.u-strasbg.fr/simbad/},  and the infrared K$_{s}$ magnitudes from 2MASS \citep{Skrutskie-2006}. The stellar mass values range from 0.09 to 0.60 $M_{\odot}$, with a mean and median values of 0.32 and 0.29 $M_{\odot}$ respectively. We note that, Gl 803, a young ($\sim 20$ Myr) M dwarf star in our sample, with a circumstellar disk \citep{Kalas-2004}, has a derived stellar mass value of 0.75, too high for a M dwarf. Therefore, the stellar mass calibration being used may not be adequate for the youngest M dwarfs.

The metallicities were first calculated with the photometric calibration provided by \citet{Neves-2012}, using stellar parallaxes, and V and K$_{s}$ magnitudes. 
%\textcolor{red}{Neves or Schlaufman calibration? It's very hard to choose one, but the Schlaufman has a higher rms = 0.11, although it also has a wider range of metallicities. BUT we can compare it with the metallicity of 451GTO, and here the Schlaufman one clearly wins.}. 
To improve on precedent photometric calibrations, we try to root the metallicity effect in the high-resolution HARPS spectra, using the measurements of the equivalent widths of the lines and features of the 26 red orders (533-690 nm region) of the HARPS spectra. The new calibration is detailed in the Appendix. We achieve a better precision with the new calibration reaching a [Fe/H] dispersion of the order of 0.08 dex. The metallicity values range from -0.88 to 0.32 dex, with a mean and median values of -0.13 and -0.11 dex respectively. We note that there is a slight offset towards lower metallicities when compared with the 582 FGK dwarfs from the HARPS-2 volume-limited sample \citep{Sousa-2011b} with mean and median values of -0.10 and -0.08 dex respectively.
%\textbf{The metallicity difference to the volume-limited and kinematically-matched sample of \citet{Schlaufman-2010}, from the Geneva Copenhagen survey \citep[GCS -- ][]{Nordstrom-2004,Holmberg-2007,Holmberg-2009}  , with a mean of -0.14 dex is, however, negligible}.

%\textcolor{red}{Note: Using the calibration of SL10 as a starting point for the feh spectroscopic calibration we have a mean of -0.12 and a median of -0.10 dex. Is the use of the spectra pushing the metallicities towards lower values, due to the molecular lines?? NO! The photometric calibrations of SL10 and Neves-2012 have even worse results (mean = -0.18, median = -0.14) BUT the SL10 puts the mean metallicity at -0.17! And the SN from GCS is -0.15.. But the end the differences are small, because the uncertainties are significative.}.

%\textcolor{red}{A table with masses and metallicities for the first 10 stars is needed in here, I think. DO for all stars and put in the end, TBD}

Table  \ref{full_table} depicts the sample used in this paper, where columns 2 and 3 list the right ascension and declination respectively, column 4 the parallaxes and their respective uncertainties, column 5 the source of the parallax, column 6 the spectral type of the M dwarf, and columns 7 and 8 the V- and K$_{s}$-band magnitudes respectively. Finally, columns 9 and 10 contain the calculated stellar mass and metallicity.

\addtocounter{table}{1}

\section{Stellar mass, metallicity, and planets from the HARPS study}
\label{relation}

In this section we use the new metallicity values (see the Appendix) as well as the stellar mass determinations from the HARPS M dwarf GTO sample to study the possible correlations of these quantities with the presence of planets. In this paper we consider Jovian hosts as stars having any planet with $M_{p} > 30 M_{\oplus}$ and Neptunian/smaller planet hosts as stars having all planets with masses below 30 $M_{\oplus}$.

%\textbf{We recall that in the M dwarf sample paper \citep{Bonfils-2011}, all planets are detected in the most massive half (resp. brightest half) of the stellar mass (resp. V-mag) distribution. This means that a stellar mass distribution may be subject to detection biases: one the one hand the reflex motion induced by the planetary companions is higher in lower mass stars, meaning a higher RV signal but on the other hand, the higher mass stars are brighter,meaning a higher signal-to-noise ratio, and thus a better RV precision. Here, we want to investigate if this trend is due to an observational bias or if it is physical in nature.}

%as shown in \citet{Bonfils-2011} paper, that 

%\textbf{We recall that in the M dwarf sample paper \citep{Bonfils-2011}, all planets are detected in the most massive half (resp. brightest half) of the stellar mass (resp. V-mag) distribution. } 

\subsection{The stellar mass-planet correlation bias}
\label{massbias}

\begin{figure}[h]
\begin{center}
\includegraphics[scale=0.45]{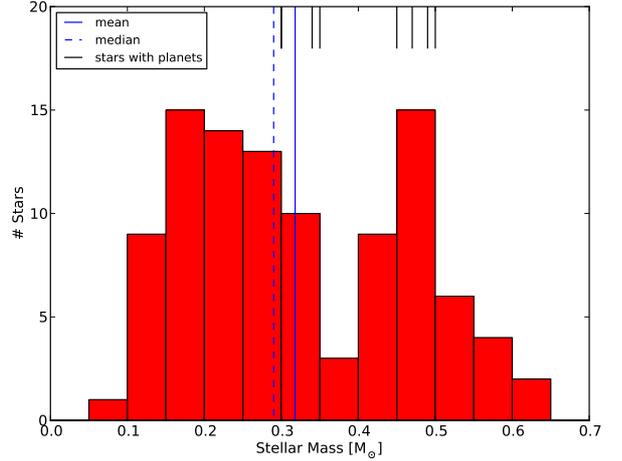}
\end{center}
\caption{Stellar mass distribution of the sample. The blue solid and dashed vertical lines represent the mean and the median of the stellar mass of the sample respectively. The black vertical lines locate the systems with planet detections.}
\label{histmass}
\end{figure}

Fig. \ref{histmass} shows the histogram of the stellar mass distribution of the whole sample. The solid blue and dashed vertical lines represent the mean and the median of the stellar mass of the sample respectively. The black vertical lines locate the systems with planet detections. 

We can see that the planet detections are all on one side of the median of our sample distribution with stellar mass (all detected planets are around the more massive stars), as previously shown by \cite{Bonfils-2011}. This is also true for the V magnitude distribution (all detected planets are around the brighter stars). Therefore, any result regarding stellar mass will be checked, because its distribution may be subject to detection biases: on the one hand the reflex motion induced by a planetary companion is higher in lower mass stars, meaning a higher radial velocity (RV) signal, but on the other hand, the lower mass stars are on average fainter, thus having higher measurement uncertainties, which makes smaller planets harder to detect.

%meaning a higher signal-to-noise ratio, and thus a better RV precision. 

%\textbf{For instance, if we take the expression of \citet{Narayan-2005} to evaluate the detection probability for low mass planets, and considering a 50\% detection rate, an orbit of 100 day period, we get, for 30 detections, and an internal uncertainty of $\sigma$ = 0.8 m/s (valid for V = 7-10 mag -- see Sect. 3 of \citet{Bonfils-2011}) a minimum detectable planetary mass of 12.13 M$_{\oplus}$ for a 1 M$_{\odot}$ solar-type star and 2.61 M$_{\oplus}$ for a 0.1 M$_{\odot}$ M dwarf. If we observe a fainter star, with V = 14 mag, we get $\sigma$ = 3.1 m/s, using $\sigma \sim 2.5^{\frac{V-10}{2}}$ from \citet{Bonfils-2011}, and we obtain a minimum detectable planetary mass of 94.75 $M_{\oplus}$ for a 1 M$_{\odot}$ star and 20.41 $M_{\oplus}$ for a 0.1 M$_{\odot}$ M dwarf.}

A lower star count in the [0.35-0.40] M$_{\odot}$ bin of Fig. \ref{histmass} is observed. To check whether this feature is real or due to a small number statistical fluctuation we did a simple monte-carlo simulation by generating 100.000 virtual samples containing 102 stars in the [0.05-0.65]  M$_{\odot}$  region, using an uniform distribution generator. Then, for each sample, we searched for a bin, in the [0.15-0.5] region, where the count difference with both adjacent bins was the same or higher than in the observed stellar mass distribution. To this end we chose a count difference of 6,7, and 8, obtaining a frequency of 10.6, 5.1, and 2.2\% respectively. We thus attribute the low number of stars with a mass between 0.35 and 0.4 M$_{\odot}$ to a small number statistical fluctuation.

%\textcolor{red}{(I wonder why Fig 4. and Fig 1. of Bonfils 2011 are a bit different..... :O)}

%as planets are easier to detect around more massive stars. In fact, the Fig. 1 of \citet{Bonfils-2011} (and Fig. \ref{histmass} of this work) shows that 

%\section{Detection limits}
%\label{dl}

To check if there is any statistically significative bias due to the detection limits in the stellar mass distribution, we will first investigate the reason why all planet detections of our sample are located in the brightest and more massive halves of the two distributions, as it was seen in Fig. \ref{histmass}, for the stellar mass. We will then confirm or deny the existence of a stellar mass-planet correlation in our sample, as shown in Table \ref{table:mass}, where we can observe a significative difference between the difference of averages and medians of giant planet and non-planet hosts.

\begin{table}[h]
\centering
\caption{Difference of averages and medians of stellar mass between planet host and non-planet host distributions. N$_{h}$ is the number of planet hosts.}
\label{table:mass}
\begin{center}
\resizebox{9cm}{!}{
\begin{tabular}{l r r }

\hline
\hline
Stellar mass & Diff. of averages & Diff. of medians \\
 & [M$_{\odot}$] & [M$_{\odot}$] \\
\hline
Full sample (N$_{h}$=8)  & 0.08 & 0.13 \\ %CHANGED to 0.08 from 0.09 for Arxiv
Jovians hosts (N$_{h}$=3) & 0.11 & 0.18 \\ %CHANGED to 0.11 from 0.12 for Arxiv
Neptunian/smaller hosts (N$_{h}$=5) & 0.07 & 0.08 \\

\hline
\hline
\end{tabular}
}
\end{center}
\end{table}

%\begin{figure}[h]
%\begin{center}
%\includegraphics[scale=0.6]{3binmass.eps}
%\end{center}
%\caption{Upper panel: Histogram of stellar mass with 3 bins for stars without planets(solid red) and stars with planets (dashed blue); Lower panel: Frequency of stars with planets.}
%\label{3binmass}
%\end{figure}

%\begin{figure}[h]
%\begin{center}
%\includegraphics[scale=0.45]{histmass.eps}
%\end{center}
%\caption{Histogram of the number of stars without planets (solid red), with Jovian planets (filled dashed blue), and with Neptunians/smaller planets (dotted black) for stellar mass. The vertical solid red, filled dashed blue, and dotted black lines above the histograms represent the mean of each distribution, respectively.}
%\label{histfullm}
%\end{figure}

In order to do this, we divided the sample into two stellar mass ranges at the median value (0.29 M$_{\odot}$). We note that we removed the star Gl803 from the sample, due to the fact that the mass for this star may have not been adequately calculated, as explained in Sect. \ref{sample}. Then, we calculated the frequency of stars with planets, using only the most massive planet in stars with multiple planets, and the frequency of planets. For both cases, we take into account the detection limits of our sample for different regions of the mass-period diagram following the procedure described in Sect. 7 of \citet{Bonfils-2011}. 

In short, for each region, we calculate the frequency $f=N_{d}/N_{\star,eff}$, where $N_{d}$ is the number of planet detections (or stars with planets), and $N_{\star,eff}$ is the number of stars whose detection limits exclude planets with similar mass and period at the 99\% confidence level. The N$_{\star,eff}$ is evaluated with Monte-Carlo sampling as described in \citet{Bonfils-2011}: we draw random mass and period within each region of study, assuming a log-uniform probability for both quantities. Then, we evaluate if the draw falls above or below the detection limit of each star. If it sits above the detection limit we include the star in the $N_{\star,eff}$. The final value of $N_{\star,eff}$ will be the average of 10.000 trials. The confidence intervals are calculated using a poissonian distribution to calculate the 1$\sigma$ gaussian-equivalent area of the probability distribution, as shown for the binomial distribution in Sect. \ref{ssec:metal}.% limit equivalent confidence intervals.%, instead of using a binomial distribution. 

The results for the two halves of the stellar mass distribution can be seen in Table \ref{fswp} for the frequency of planet-hosts (N=8), and in Table \ref{fpl} for the occurrence of planets (N=14). We observe that, in the planet-host case, all values between the upper limits for M$_{\star} \le 0.29M_{\odot}$ and the frequency values  for M$_{\star} > 0.29M_{\odot}$ are compatible with each other for all regions of planetary mass and period, except in the three regions with period between 10 and $10^{4}$ days, and mass between 1 and 10 M$_{\oplus}$, where we cannot compare the values due to a low N$_{eff}$ number. We observe the same regarding the results of the occurrence of planets. 

\begin{table*}[th!]
\begin{center}
\caption{a) Upper limits for the occurrence of planet-hosts for M$_{\star} \leq 0.29$ $M_{\odot}$ (N$_{\star}$=52); b) Frequencies and upper limits for the occurrence of planet-hosts for M$_{\star}>  0.29$ $M_{\odot}$ (N$_{\star}$=49). Multi-planet hosts are characterized by their most massive planet. }
\label{fswp}
\subtable[]{
\tiny
\resizebox{8.5cm}{!}{

\begin{tabular}{l | c c c c}
\hline
\hline

                &  \multicolumn{4}{c}{Period} \\
$m\sin{i}$ & \multicolumn{4}{c}{[day]} \\

\multicolumn{1}{c |}{[M$_{\oplus}$]}  & $1-10$ & $10 - 10^{2}$ & $10^{2} - 10^{3}$ & $10^{3} - 10^{4}$ \\
\hline
$10^3 - 10^4$ & $N_{d}=0$ & $N_{d}=0$ & $N_{d}=0$ & $N_{d}=0$ \\
              & $N_{eff} = 47.51$ & $N_{eff} = 46.85$ & $N_{eff} = 45.74$ & $N_{eff} = 42.67$ \\
              & $f<0.02(1\sigma)$ & $f<0.02(1\sigma)$ & $f<0.02(1\sigma)$ & $f<0.03(1\sigma)$ \\
$10^2 - 10^3$ & $N_{d}=0$ & $N_{d}=0$ & $N_{d}=0$ & $N_{d}=0$ \\
              & $N_{eff} = 44.11$ & $N_{eff} = 41.19$ & $N_{eff} = 36.31$ & $N_{eff} = 24.39$ \\
              & $f<0.03(1\sigma)$ & $f<0.03(1\sigma)$ & $f<0.03(1\sigma)$ & $f<0.05(1\sigma)$ \\
$10 - 10^2$ & $N_{d}=0$ & $N_{d}=0$ & $N_{d}=0$ & $N_{d}=0$ \\
              & $N_{eff} = 28.56$ & $N_{eff} = 18.86$ & $N_{eff} = 9.90$ & $N_{eff} = 3.43$ \\
              & $f<0.04(1\sigma)$ & $f<0.06(1\sigma)$ & $f<0.12(1\sigma)$ & $f<0.31(1\sigma)$ \\
$1 - 10$ & $N_{d}=0$ & $N_{d}=0$ & $N_{d}=0$ & $N_{d}=0$ \\
              & $N_{eff} = 3.90$ & $N_{eff} = 1.45$ & $N_{eff} = 0.46$ & $N_{eff} = 0.01$ \\
              & $f<0.28(1\sigma)$ & $ - $ & $ - $ & $ - $ \\ [1pt]

\hline
\hline
\end{tabular}}}
\subtable[]{
\tiny
\resizebox{8.35cm}{!}{
\begin{tabular}{l | c c c c}

%\caption{Frequencies and upper limits for the occurrence of stars with planets for M$_{\star} > 0.29$ $M_{\odot}$ (N$_{\star}$=49). Multiplanetary systems are characterized by their most massive planet.}

\hline
\hline

                &  \multicolumn{4}{c}{Period} \\
$m\sin{i}$ & \multicolumn{4}{c}{[day]} \\

\multicolumn{1}{c |}{[M$_{\oplus}$]}  & $1-10$ & $10 - 10^{2}$ & $10^{2} - 10^{3}$ & $10^{3} - 10^{4}$ \\
\hline
$10^3 - 10^4$ & $N_{d}=0$ & $N_{d}=0$ & $N_{d}=0$ & $N_{d}=0$ \\
              & $N_{eff} = 48.93$ & $N_{eff} = 48.73$ & $N_{eff} = 48.34$ & $N_{eff} = 47.24$ \\
              & $f<0.02(1\sigma)$ & $f<0.02(1\sigma)$ & $f<0.02(1\sigma)$ & $f<0.02(1\sigma)$ \\
$10^2 - 10^3$ & $N_{d}=0$ & $N_{d}=1$ & $N_{d}=0$ & $N_{d}=2$ \\
              & $N_{eff} = 47.79$ & $N_{eff} = 47.03$ & $N_{eff} = 44.74$ & $N_{eff} = 34.66$ \\
              & $f<0.02(1\sigma)$ & $f=0.02_{-0.01}^{+0.05}$ & $f<0.03(1\sigma)$ & $f=0.06_{-0.02}^{+0.08}$ \\
$10 - 10^2$ & $N_{d}=2$ & $N_{d}=0$ & $N_{d}=0$ & $N_{d}=0$ \\
              & $N_{eff} = 40.26$ & $N_{eff} = 31.78$ & $N_{eff} = 19.98$ & $N_{eff} = 7.18$ \\
              & $f=0.05_{-0.02}^{+0.07}$ & $f<0.04(1\sigma)$ & $f<0.06(1\sigma)$ & $f<0.16(1\sigma)$ \\
$1 - 10$ & $N_{d}=3$ & $N_{d}=0$ & $N_{d}=0$ & $N_{d}=0$ \\
              & $N_{eff} = 9.44$ & $N_{eff} = 3.89$ & $N_{eff} = 0.98$ & $N_{eff} = 0.10$ \\
              & $f=0.32_{-0.10}^{+0.31}$ & $f < 0.28 (1\sigma)$ & $ - $ & $ - $ \\ [1pt]
\hline
\hline
\end{tabular}}}
\end{center}
\end{table*}

\begin{table*}[ht!]
%\centering
\begin{center}
\caption{a) Upper limits for the occurrence of planets for M$_{\star} \leq 0.29$ $M_{\odot}$ (N$_{\star}$=52); b) Frequencies and upper limits for the occurrence of planets for M$_{\star} > 0.29$ $M_{\odot}$ (N$_{\star}$=49).}
\label{fpl}
\subtable[]{
\tiny
\resizebox{8.5cm}{!}{
\begin{tabular}{l | c c c c}

\hline
\hline

                &  \multicolumn{4}{c}{Period} \\
$m\sin{i}$ & \multicolumn{4}{c}{[day]} \\

\multicolumn{1}{c |}{[M$_{\oplus}$]}  & $1-10$ & $10 - 10^{2}$ & $10^{2} - 10^{3}$ & $10^{3} - 10^{4}$ \\
\hline
$10^3 - 10^4$ & $N_{d}=0$ & $N_{d}=0$ & $N_{d}=0$ & $N_{d}=0$ \\
              & $N_{eff} = 47.51$ & $N_{eff} = 46.85$ & $N_{eff} = 45.74$ & $N_{eff} = 42.70$ \\
              & $f<0.02(1\sigma)$ & $f<0.02(1\sigma)$ & $f<0.02(1\sigma)$ & $f<0.03(1\sigma)$ \\
$10^2 - 10^3$ & $N_{d}=0$ & $N_{d}=0$ & $N_{d}=0$ & $N_{d}=0$ \\
              & $N_{eff} = 44.13$ & $N_{eff} = 41.24$ & $N_{eff} = 36.45$ & $N_{eff} = 24.63$ \\
              & $f<0.03(1\sigma)$ & $f<0.03(1\sigma)$ & $f<0.03(1\sigma)$ & $f<0.05(1\sigma)$ \\
$10 - 10^2$ & $N_{d}=0$ & $N_{d}=0$ & $N_{d}=0$ & $N_{d}=0$ \\
              & $N_{eff} = 28.51$ & $N_{eff} = 18.84$ & $N_{eff} = 9.89$ & $N_{eff} = 3.46$ \\
              & $f<0.04(1\sigma)$ & $f<0.06(1\sigma)$ & $f<0.12(1\sigma)$ & $f<0.31(1\sigma)$ \\
$1 - 10$ & $N_{d}=0$ & $N_{d}=0$ & $N_{d}=0$ & $N_{d}=0$ \\
              & $N_{eff} = 3.92$ & $N_{eff} = 1.47$ & $N_{eff} = 0.47$ & $N_{eff} = 0.01$ \\
              & $f<0.28(1\sigma)$ & $ - $ & $ - $ & $ - $ \\ [1pt]
\hline
\hline
\end{tabular}}}
\subtable[][]{
\tiny
\resizebox{8.35cm}{!}{
\begin{tabular}{l | c c c c}

\hline
\hline

                &  \multicolumn{4}{c}{Period} \\
$m\sin{i}$ & \multicolumn{4}{c}{[day]} \\

\multicolumn{1}{c |}{[M$_{\oplus}$]}  & $1-10$ & $10 - 10^{2}$ & $10^{2} - 10^{3}$ & $10^{3} - 10^{4}$ \\
\hline
$10^3 - 10^4$ & $N_{d}=0$ & $N_{d}=0$ & $N_{d}=0$ & $N_{d}=0$ \\
              & $N_{eff} = 48.92$ & $N_{eff} = 48.71$ & $N_{eff} = 48.34$ & $N_{eff} = 47.21$ \\
              & $f<0.02(1\sigma)$ & $f<0.02(1\sigma)$ & $f<0.02(1\sigma)$ & $f<0.02(1\sigma)$ \\
$10^2 - 10^3$ & $N_{d}=0$ & $N_{d}=2$ & $N_{d}=0$ & $N_{d}=2$ \\
              & $N_{eff} = 47.78$ & $N_{eff} = 47.02$ & $N_{eff} = 44.65$ & $N_{eff} = 34.48$ \\
              & $f<0.02(1\sigma)$ & $f=0.04_{-0.01}^{+0.06}$ & $f<0.03(1\sigma)$ & $f=0.06_{-0.02}^{+0.08}$ \\
$10 - 10^2$ & $N_{d}=2$ & $N_{d}=0$ & $N_{d}=1$ & $N_{d}=0$ \\
              & $N_{eff} = 40.23$ & $N_{eff} = 31.60$ & $N_{eff} = 19.85$ & $N_{eff} = 7.23$ \\
              & $f=0.05_{-0.02}^{+0.07}$ & $f<0.04(1\sigma)$ & $f=0.05_{-0.01}^{+0.12}$ & $f<0.16(1\sigma)$ \\
$1 - 10$ & $N_{d}=5$ & $N_{d}=3$ & $N_{d}=0$ & $N_{d}=0$ \\
              & $N_{eff} = 9.46$ & $N_{eff} = 3.90$ & $N_{eff} = 0.99$ & $N_{eff} = 0.10$ \\
              & $f=0.53_{-0.15}^{+0.36}$ & $f=0.77_{-0.23}^{+0.75}$ & $ - $ & $ - $ \\ [1pt]

\hline
\hline
\end{tabular}}}

\end{center}
\end{table*}

The fact that we do not observe a statistically significative ($> 2\sigma$) difference in any region of the mass-period diagram between the two stellar mass sub-samples indicate that the observed accumulation of planet hosts in the higher half of the stellar mass distribution is due to a stellar mass detection bias. Therefore, we will not study the stellar mass-planet relation any further for our HARPS sample.

We got similar results for the V magnitude distribution, as the brightness and stellar mass have similar effects regarding the precision of the RV measurements.

%We need more detections to check if a real stellar mass trend exists or not.

%Other than that, we cannot conclude anything regarding whether the observed stellar mass-planet dependence in Fig \ref{3binmass} is true or if it is just an observational bias towards brighter/ more massive stars.

\subsection{The metallicity-planet correlation}
\label{ssec:metal}

\begin{figure}[h]
\begin{center}
\includegraphics[scale=0.45]{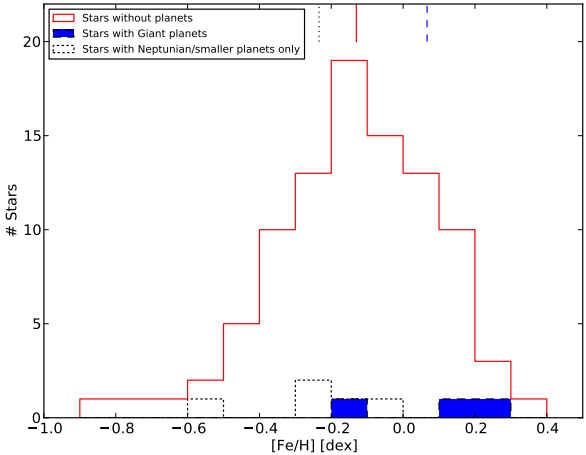}
\end{center}
\caption{Histograms of stars without planets (solid red), with Jovian planets (filled dashed blue), and with Neptunians/smaller planets (dotted black) for metallicity. The vertical solid red, filled dashed blue, and dotted black lines above the histograms represent the mean of the [Fe/H] distribution.}
\label{histfull}
\end{figure}

Figure \ref{histfull} shows the histogram of metallicity of our sample. The solid red histogram represent the stars without planets, while the filled dashed blue histogram the stars with Jovians planets, and the dotted black histogram the star with Neptunians/smaller planets only. The vertical solid red, dashed blue, and dotted black lines above each histogram depict the value of the mean of the distribution. We note here that we assume that metallicity is not influenced by detection biases, due to the fact that we are using a volume-limited sample.

%We note that there is a lower count in the 0.35-0.40 M$_{\odot}$ bin. We attribute this to statistical fluctuations due to the small number of each bin. \textcolor{red}{what else???}.

We can observe in Table \ref{feh:planets} that the difference of the averages (medians resp.) of the full sample between planet and non-planet host distributions is small (0.01 and -0.07 dex, respectively). %For stellar mass, there is a 0.09 $M_{\odot}$ (0.13 $M_{\odot}$ resp.) difference between the averages (medians resp.) of planet and non-planet host histograms.

\begin{table}[h]
\centering
\caption{Difference of averages and medians of [Fe/H] between planet host and non-planet host distributions. N$_{h}$ is the number of planet hosts.}
\label{feh:planets}
\begin{center}
\resizebox{9cm}{!}{
\begin{tabular}{l r r r }

\hline
\hline
[Fe/H] & Diff. of averages & Diff. of medians & KS test \\
 & [dex] & [dex] & \\
 \hline
Full sample (N$_{h}$=8) & 0.01 & -0.07 & 0.8151\\
Jovians hosts (N$_{h}$=3) & 0.20 & 0.26 & 0.1949\\
Neptunian/smaller hosts (N$_{h}$=5) & -0.10 & -0.11 & 0.3530 \\ %CHANGE the median from -0.10 to -0.11 for Arxiv
\hline
\hline
\end{tabular}
}
\end{center}
\end{table}

If we only take into account the three planet host stars with Jupiter-type planets, the difference of the averages and the medians of the [Fe/H] between stars with and without planets is higher (0.20 and 0.26 dex respectively). On the other hand, if we remove the 3 systems with Jupiters, we obtain -0.10 dex for the means and -0.11 dex for the medians. The correlation we find between [Fe/H] and planet occurrence agrees with previous studies focused on giant planets around M dwarfs \citep[e.g.][]{Bonfils-2007,Johnson-2009,Johnson-2010,Schlaufman-2010,Rojas-Ayala-2010,Rojas-Ayala-2012,Terrien-2012}. We confirm also, with better statistics, that such correlation is vanishing for Neptunian and smaller planet hosts \citep[e.g.][]{Rojas-Ayala-2012,Terrien-2012}. In fact our result hints at a anti-correlation between [Fe/H] and planets though the difference (-0.10 dex) is at the limit of our measurement precision. Despite that, the results hint a different type of planet formation mechanism for giant and Neptunian/Super Earth-type planets \citep[e.g.][]{Mordasini-2012}.%(\textcolor{red}{check/add reference}).

%\citep[e.g.][]{Sousa-2008,Bouchy-2009,Sousa-2011b, Mayor-2011} and M dwarfs , where such correlation seems to vanish. 

%Regarding stellar mass, we observe a slight increase in the difference of averages (medians resp.) between planet and non-planet host distributions (0.12 and 0.18 dex respectively). For hosts having smaller planets only, a mean of 0.07 and a median of 0.08 dex is observed.

% \textcolor{red}{histogram picture for the jovians and neptunians here?}

We performed a Kolmogorov-Smirnov (KS) test to check the probability of the sub-samples of stars with and without planets of belonging to the same parent distribution. All KS tests show that we cannot discard the possibility that the three sub-samples with planets belong to the same distribution of the stars without planets. We obtain a value of 0.195 for the Jovians hosts, but we do not have enough hosts (N=3) to calculate the KS test properly.

%We did not calculate the KS p-values for the Neptunian and Jovian host sub-samples because they do not have enough members to calculate the KS test properly.

%However, this result is coherent with the [Fe/H] results for Neptunian type FGK host stars where the planet-metallicity relation seems to vanish \citep. . %Moreover, we must note that we only have 8 stars with planets within our sample, which should make us cautious regarding the results of the K-S test. 

% thus not agreeing with previous results . 

%For stellar mass, there is a low 1\% probability that the stars with and without planets are drawn from the same distribution, hinting a real difference between the samples, in accordance with studies using much wider spectral ranges, with
%M dwarfs \citep{Johnson-2007,Johnson-2010} as well as for 
%AFGKM dwarf samples \citep{Laws-2003,Lovis-2007,Johnson-2010}. %\textcolor{red}{check these references.........!!! and put some model references.} %\textcolor{red}{It is interesting to see that, if we divide the planet host sample into one subgroup having only Jupiter-type planets (N=3) and into another subgroup having Neptunian and smaller type planets (N=5) we obtain similar values for the difference between planet hosts and non-planet hosts (. This is a hint that 

%For the same reason we did not study the frequency of giant planets around M dwarfs, as we only have three Jupiters type planets in two systems. 

%Nevertheless, if we separate the planet-host sample into two subgroups, with  divided by the sum of the planetary masses,

\begin{figure}[h]
\begin{center}
\includegraphics[scale=0.6]{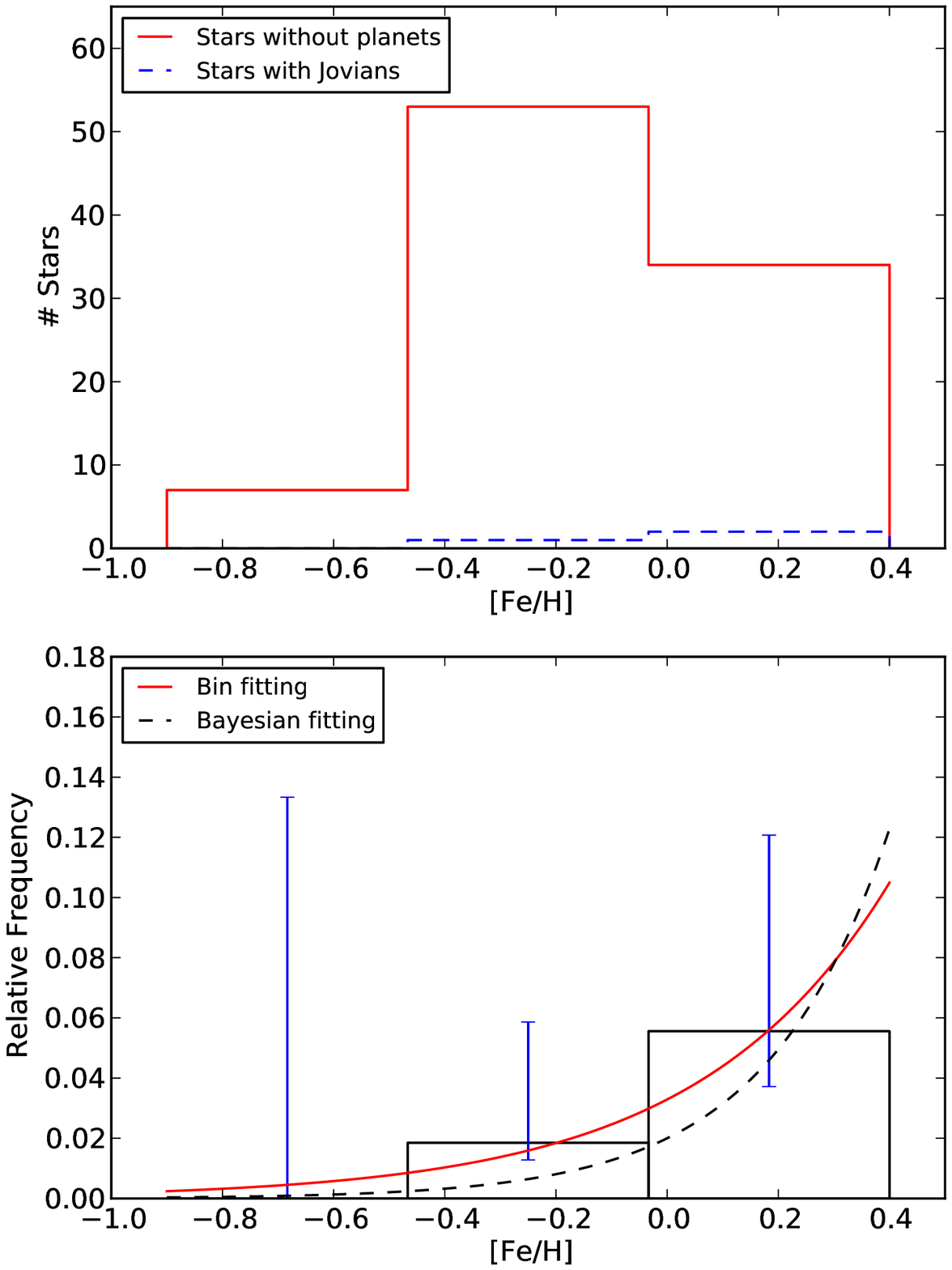}
\end{center}
\caption{Upper panel: Histogram of metallicity with 3 bins for stars without planets (solid red) and stars with Giant planets (dashed blue); Lower panel: Frequency of stars with Giant planets.}
\label{3binfehj}
\end{figure}

\begin{figure}[h]
\begin{center}
\includegraphics[scale=0.6]{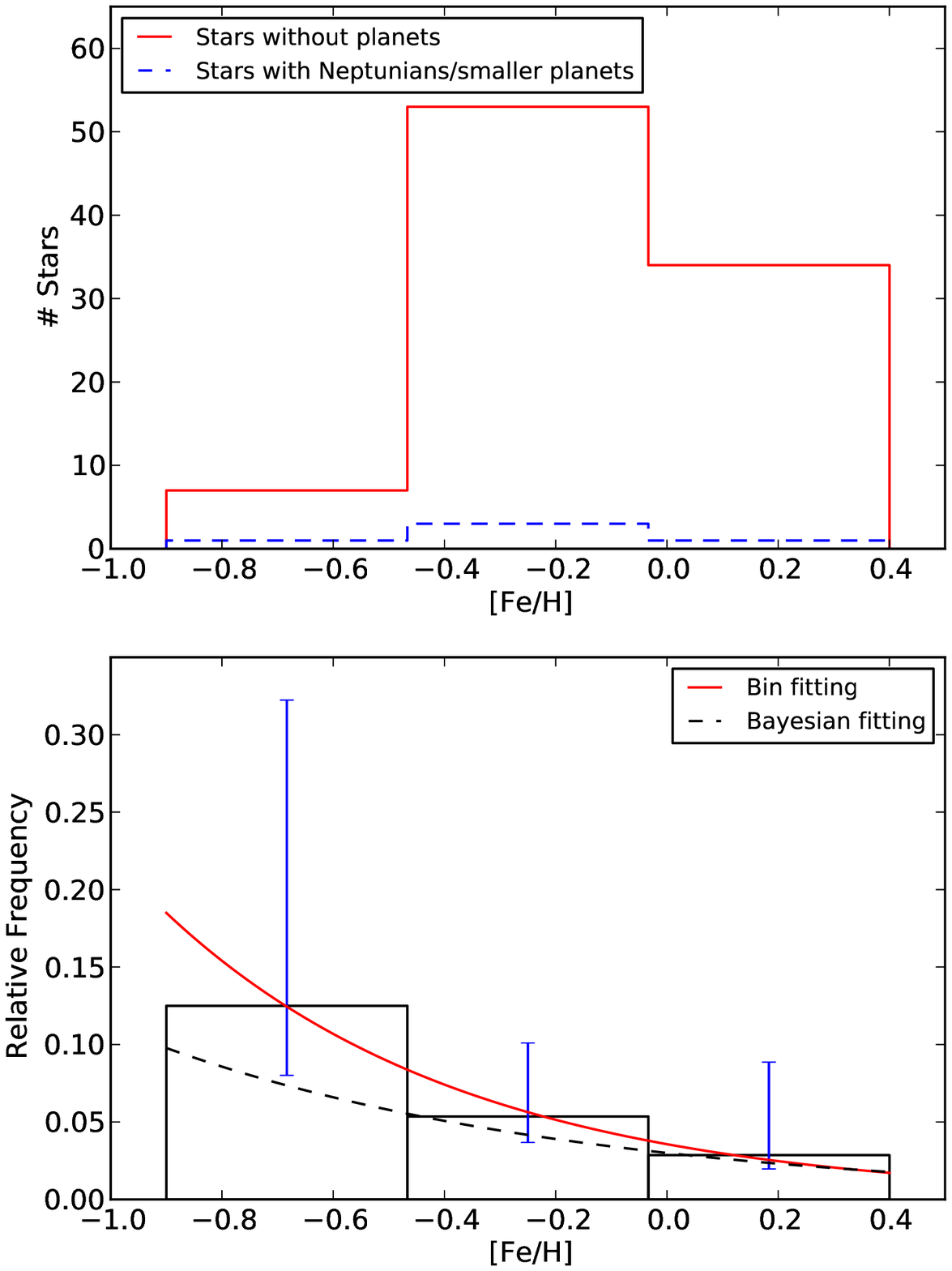}
\end{center}
\caption{Upper panel: Histogram of metallicity with 3 bins for stars without planets (solid red) and stars with Neptunians and smaller planets (dashed blue); Lower panel: Frequency of stars with Neptunians and smaller planets.}
\label{3binfehn}
\end{figure}

In order to explore the star-planet relation further, we divided the metallicity range in three bins and performed a frequency analysis for Jovian hosts and Neptunian/smaller planet hosts separately, as shown in Figs. \ref{3binfehj} and \ref{3binfehn}. %Regarding stellar mass, we only consider the full sample here because, as we will see in Sect. \ref{dl}, the correlation that is found in Fig. \ref{3binmass} is due to a detection bias. 
The upper panels of all figures are the same as in Fig. \ref{histfull}, but this time with only three bins. 

\begin{figure}[h]
\begin{center}
\includegraphics[scale=0.45]{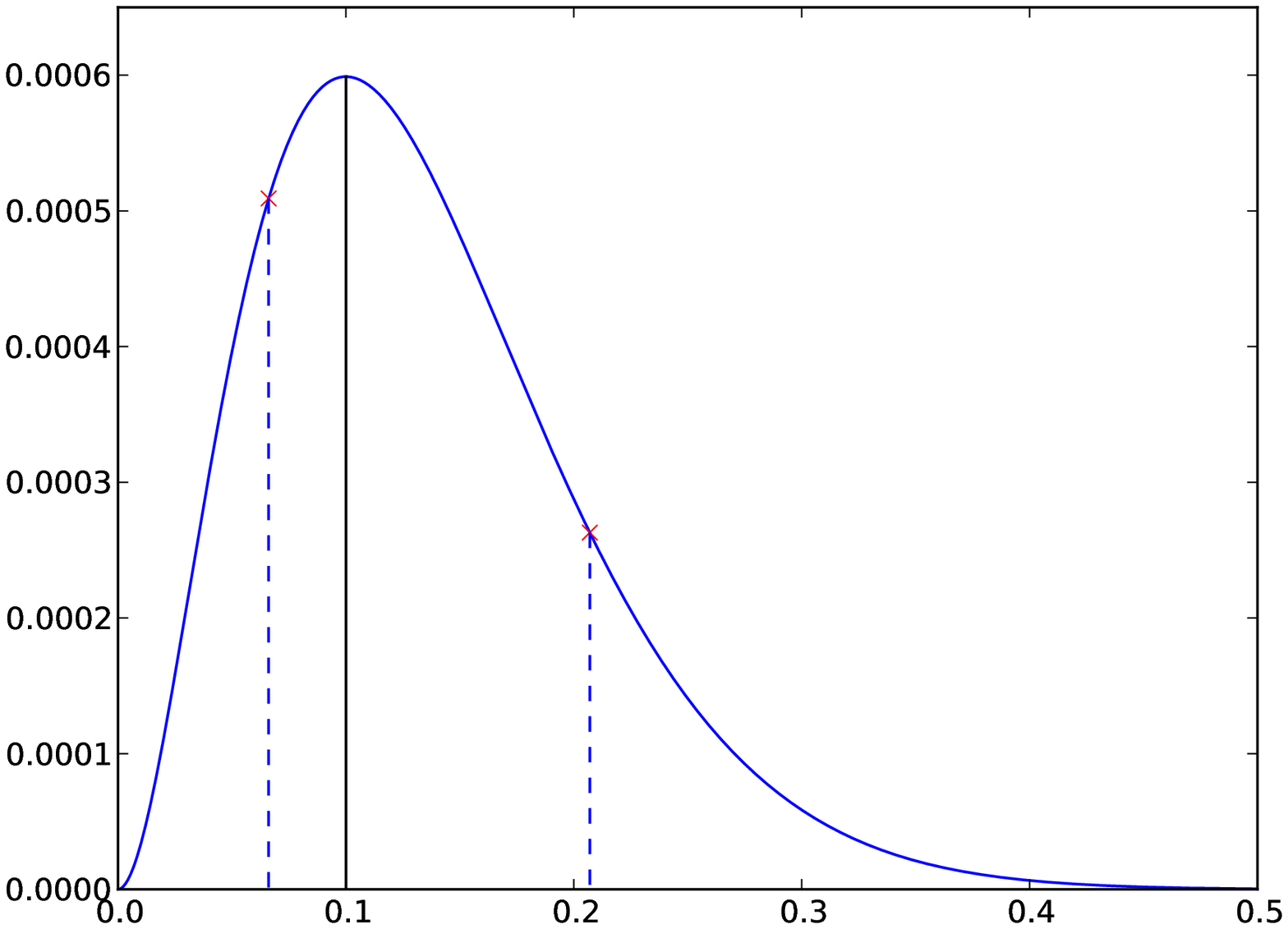}
\end{center}
\caption{Normalized binomial probability distribution function for $n=2$, $N=20$, and $f_{p} = 0.1$.The solid vertical line depicts the observed frequency. The dashed lines show the 68.2\% ($1\sigma$) limits around the maximum of the function.}
\label{binompdf}
\end{figure}

The lower panels depict the relative frequency of the stars with planets. The solid red line corresponds to a direct least squares bin fitting, while the dashed black line is a bayesian bin-independent parametric fitting, explained in Sect. \ref{bayes}. Both fits use the functional form $f=C10^{\alpha[Fe/H]}$, following previous works for FGK dwarfs \citep[][]{Valenti-2005,Udry-2007,Sousa-2011b}. %and will be discussed in detail in Sect. \ref{bayes}. 
The coefficients $C$ and $\alpha$ of both methods and respective uncertainties are shown in Table \ref{param:table}. The errors in the frequency of each bin are calculated using the binomial distribution, 

\begin{equation}
P(f_{p},n,N) = \frac{N!}{n!(N-n)!}f^{n}_{p}(1-f_{p})^{N-n},
\label {binom}
\end{equation}
following the procedure outlined in, e.g., \citet[][]{Burgasser-2003,McCarthy-2004, Endl-2006}, and \citet{Sozzetti-2009}. In short we calculate how many $n$ detections we have in a bin of size $N$, as a function of the planet frequency $f_{p}$, of each bin. The upper errors, lower errors and upper limits of each bin are calculated by measuring the 68.2\% of the integrated area around the peak of the binomial probability distribution function, that corresponds to the 1$\sigma$ limit for a gaussian distribution. An example is shown in Fig. \ref{binompdf}, depicting a normalized binomial probability distribution function with $n = 2$, $N = 20$, and $f_{p} = 0.1$. 

From Figs. \ref{3binfehj} and \ref{3binfehn} it can be observed that there is a small statistical difference between the frequency bins for both Jovian-hosts and Neptunian and smaller planet hosts, as the uncertainties of each bin are high. The first bin of Fig. \ref{3binfehj} ([-0.9,0.5] dex) has an upper limit of 13.3\%, with no planet detection, while the second and third bins  ([-0.5,-0.1] and [-0.1,0.3] dex, resp.) have values of 1.9\% and 5.6\% respectively. Regarding Fig. \ref{3binfehn}, we observe the upper limit of 12.5, and the frequencies of  5.3, and 2.8\% for the same bins.

We can observe a correlation with [Fe/H]  for Jovian hosts and a hint of an anti-correlation for Neptunian and smaller planets only hosts. Interestingly, the later anti-correlation for smaller planet hosts is predicted by recent studies using core-accretion models \citep{Mordasini-2012}, but we note that we only consider Neptunian hosts as star with Neptunians and smaller planets only: if a multi-planet system has a Jovian and one or more smaller planets, for instance, we count the system as being a Jupiter host, not a Neptunian-host. Therefore, it is expected that the number of Neptunians and smaller planets will be higher at lower metallicities. %Regardless, we cannot conclude anything, due to the high uncertainties of the frequencies, related to the low number of planet hosts in our sample. 

\subsection{Bayesian approach}
\label{bayes}

To test the metallicity results we performed a parametric and bin-independent fitting of the data based on bayesian inference. We followed the \citet{Johnson-2010} approach, using two functional forms for the planet frequency, $f_{p1} = C$ and $f_{p2} = C10^{\alpha [Fe/H]}$, and choosing uniformly distributed priors for the parameters C and $\alpha$. The choice of a power law for the functional form was based on previous works of [Fe/H] of FGK dwarfs \citep[][]{Valenti-2005,Udry-2007,Sousa-2011b}. 

Table \ref{param:table} summarizes and compares the results of the Bayesian fitting to the ones obtained with the bin fitting. Column 1 shows the functional forms used and respective parameters, column 2 the uniform prior range, column 3 the most likely value for the fit parameters, along with the $1\sigma$ gaussian uncertainties and column 4 the fit parameters of the least squares bin fitting.

\begin{table}[h]
\caption{Parameters of the bayesian and fit from binning models for the HARPS sample.} %\textcolor{red}{To be updated!}}
\label{param:table}
\resizebox{9cm}{!}{
\begin{tabular}{l | c c c}

\hline
\hline
Parameters  & Uniform  & most likely  & fit from   \\
for Jovian hosts      & Prior       & value & binning \\
\hline
$f_{p1} = C$ & &  & \\
C & (0.01,0.30) & $0.03\pm0.02$ & $0.02\pm$0.02 \\
%\hline
$f_{p2} = C10^{\alpha[Fe/H]}$ & & & \\
C & (0.01,0.30) & $0.02\pm0.02$ & $0.03\pm0.01$ \\
$\alpha$ & (-1.0,4.0) & $1.97\pm1.25$ & $1.26\pm0.30$\\
\hline
\hline
Parameters  & Uniform  & most likely  & fit from   \\
for Neptunian hosts      & Prior       & value & binning \\
\hline
$f_{p1} = C$ & &  & \\
C & (0.01,0.30) & $0.05\pm0.02$ & $0.07\pm0.04$ \\
%\hline
$f_{p2} = C10^{\alpha[Fe/H]}$ & & & \\
C & (0.01,0.30) & $0.03\pm0.02$ & $0.04\pm0.01$  \\
$\alpha$ & (-4.0,1.0) & $-0.57\pm0.71$ & $-0.79\pm0.06$\\
\hline
\hline

\end{tabular}
}
\end{table}

%We note that the $\alpha$ value of the $f_{p} = C10^{\alpha [Fe/H]}$ functional form has large uncertainties. The $\alpha$ value here can easily accommodate both positive or negative values. 

From Table \ref{param:table} we can see that the Bayesian fit values are, in general, compatible with the bin fitting values. However, we observe that the $\alpha$ values obtained for the planet-host frequencies with the Bayesian method are higher than the same values using the bin fitting. This translates into a higher Giant-host frequency values with [Fe/H] and a lower Neptunian/smaller planet host frequencies as a function of metallicity. We also note that the $\alpha$ values calculated by the Bayesian method have large uncertainties in both scenarios. In the case of Neptunian-hosts, the $\alpha$ value can easily accommodate both positive or negative values.

\subsection{Comparison with the California Planet Survey late-K and M-type dwarf sample}

Our aim here is to compare our results to a similar sample regarding the difference between planet hosts and non-planet hosts only. The California Planet Survey (CPS) late-K and M-type dwarf sample \citep[][]{Rauscher-2006,Johnson-2010b} was chosen for this goal. It is a 152 star sample where 18 planets (7 Jovians and 11 Neptunian/smaller planets) are already detected around 11 hosts. The metallicities and stellar masses were calculated using the \citet{Johnson-2009} and the \citet{Delfosse-2000} calibration, respectively. We note that the \citet{Johnson-2009} [Fe/H] calibration has a dispersion around $\sim$ 0.2 dex and a systematic offset towards higher [Fe/H], as shown in \citet{Neves-2012}. The offset amounts to 0.13 dex when we compare the [Fe/H] of the CPS sample computed from the \citet{Johnson-2009} calibration with the \citet{Neves-2012} calibration.

Table  \ref{full_table_CPS} depicts the CPS sample used in this paper, where columns 2 and 3 list the right ascension and declination respectively, column 4 the parallaxes and their respective uncertainties, column 5 the source of the parallax, column 6 the spectral type of the star, and columns 7 and 8 the V- and K$_{s}$-band magnitudes respectively. Column 9 lists the stellar mass. Finally, columns 10 and 11 contain the calculated metallicity using the \citet{Johnson-2009} and the \citet{Neves-2012} photometric calibrations respectively.

\addtocounter{table}{1}

We calculated the difference of averages and medians between planet hosts and non-planet hosts in the same way as we did for our sample, as shown in Table \ref{feh:planets}. Table \ref{planets:keck} shows the results. For metallicity, we observe a much higher difference of averages and medians when compared to our sample, but as we noted before there is an offset when calculating the metallicity with different calibrations. %As shown by \citet{Neves-2012} the \citet{Johnson-2009} calibration has an offset towards a higher metallicity.} %but this only reflects the higher number of Jovian hosts of the CPS sample. 
The difference of averages and medians for Jupiter-type planets is higher than in our sample but is compatible with our results. For Neptunian-type hosts the difference of averages and medians are indistinguishable from the non-planet host sample. 

We also performed a KS test for [Fe/H] between the three planet-host subsamples and the stars without planets, taking advantage of the higher number of stars with planets of the CPS sample, as shown in the forth column of Table \ref{planets:keck}. It can be seen that there is a very low probability ($\sim$0.2\%) that the Jovian hosts and the stars without planets belong to the same distribution. For the case of Neptunian-hosts, however, the KS p-value is high ($\sim$98\%). Again, this result is expected from previous works on FGK dwarfs \citep[e.g.][]{Sousa-2011b} and M dwarfs \citep[e.g.][]{Rojas-Ayala-2012}.

Regarding stellar mass, we do not see any trend. The difference of averages and medians between planet hosts and non-planet hosts is negligible. This result agrees with the findings of the HARPS sample as the trend we observe with stellar mass is biased.

\begin{table}[h]
\centering
\caption{Difference of averages and medians between planet host and non-planet host distributions for the CPS late-K and M-type dwarf sample.}
\label{planets:keck}
\begin{center}
\resizebox{9cm}{!}{
\begin{tabular}{l r r r}

\hline
\hline
[Fe/H] & Diff. of averages & Diff. of medians & KS test \\
 & [dex] & [dex] & \\
\hline
Full sample (N$_{h}$=11) & 0.18 & 0.21 & 0.0357\\
Jovians hosts (N$_{h}$=6) & 0.37 & 0.33 & 0.0017\\
Neptunian/smaller hosts (N$_{h}$=5) & -0.03 & -0.05 & 0.9827 \\
\hline
Stellar mass & Diff. of averages & Diff. of medians \\
 & [M$_{\odot}$] & [M$_{\odot}$] \\
Full sample (N$_{h}$=11)  & -0.04 & -0.01 & \\
Jovians hosts (N$_{h}$=6) & -0.03 & -0.05 \\
Neptunian/smaller hosts (N$_{h}$=5) & -0.04 & 0.00 \\

\hline
\hline
\end{tabular}
}
\end{center}
\end{table}

%Figure \ref{3feh:keck} shows, it its upper panel, the histograms of non-planet hosts (solid red line) and stars with planets (dashed blue line) of the CPS sample, similar to Fig. \ref{3binfeh} of our sample. The lower panel depicts the frequency of planets of each bin. It is interesting to see that with the higher planet-host count we may start to see a build-up of Neptunian and smaller planet host stars at lower metallicities and, as we go towards higher [Fe/H] values we observe the well established [Fe/H]-Giant planet host relation \citep[e.g.][]{Santos-2004b,Fischer-2005}. We must be cautious, however, as the number of planet hosts is still small. The observed frequencies, ranging from 20.0\% to 5.1\% are compatible with the results obtained with our sample.

%\begin{figure}[h]
%\begin{center}
%\includegraphics[scale=0.6]{3binfeh_keck.eps}
%\end{center}
%\caption{Upper panel: Histogram of metallicity with 3 bins for stars without planets(solid red) and stars with planets (dashed blue) for the CPS sample; Lower panel: Frequency of stars with planets.}
%\label{3feh:keck}
%\end{figure}

\section{ Metallicity-planet relation from the HARPS+CPS joined sample}

To improve our statistics and study the planet-metallicity correlation in more detail, we joined our HARPS sample with the CPS M dwarf sample. The [Fe/H] for the CPS sample was recalculated with the \citet{Neves-2012} calibration, which has the same scale and accuracy of our new calibration, shown in the Appendix. We kept the values of the [Fe/H] using our new spectroscopic calibration for the 49 stars in common. The joined sample has 205 stars, with 13 stars hosting 20 planets. Seven hosts have Jovian-type planets around them while six of them only have Neptunians and smaller planets. 

Table \ref{planets:joined} shows the results for the joined sample, and is similar to Table \ref{planets:keck}. We did not calculate the correlation between planets occurency and stellar mass, because as discussed in Sect. \ref{massbias} such relation is biased. The joined sample results are similar to both our sample and the CPS sample: the difference of averages and medians between Jovian hosts and non-planet hosts show a correlation with [Fe/H], while the same quantities for Neptunians and smaller hosts do not show this trend. The tentative hint of an anti-correlation withÊ [Fe/H] for the Neptunians/smaller hosts of the HARPS sample, in Table \ref{feh:planets} is observed but is smaller than the one observed for the HARPS sample. However, we must note that the CPS sample is not as sensitive as the HARPS sample in the detection of Neptunian and smaller planets. Therefore we consider that in this paper the reference is the HARPS sample regarding the Neptunian-host metallicity relation.

\begin{table}[h]
\centering
\caption{Difference of averages and medians between planet host and non-planet host distributions for the joined sample.}
\label{planets:joined}
\begin{center}
\resizebox{9cm}{!}{
\begin{tabular}{l r r r}

\hline
\hline
[Fe/H] & Diff. of averages & Diff. of medians & KS test \\
 & [dex] & [dex] & \\
\hline
Full sample (N$_{h}$=13) & 0.08 & 0.12 & 0.2380\\
Jovians hosts (N$_{h}$=7) & 0.20 & 0.19 & 0.0159\\
Neptunian/smaller hosts (N$_{h}$=6) & -0.05 & -0.06 & 0.8006 \\
\hline
\hline
\end{tabular}
}
\end{center}
\end{table}

The KS test results are similar to the ones performed for the CPS sample, in Table \ref{planets:keck}. However we must note the higher value in the case of the Jovian hosts, just above the 1\% p-value.

\begin{figure}[h]
\begin{center}
\includegraphics[scale=0.6]{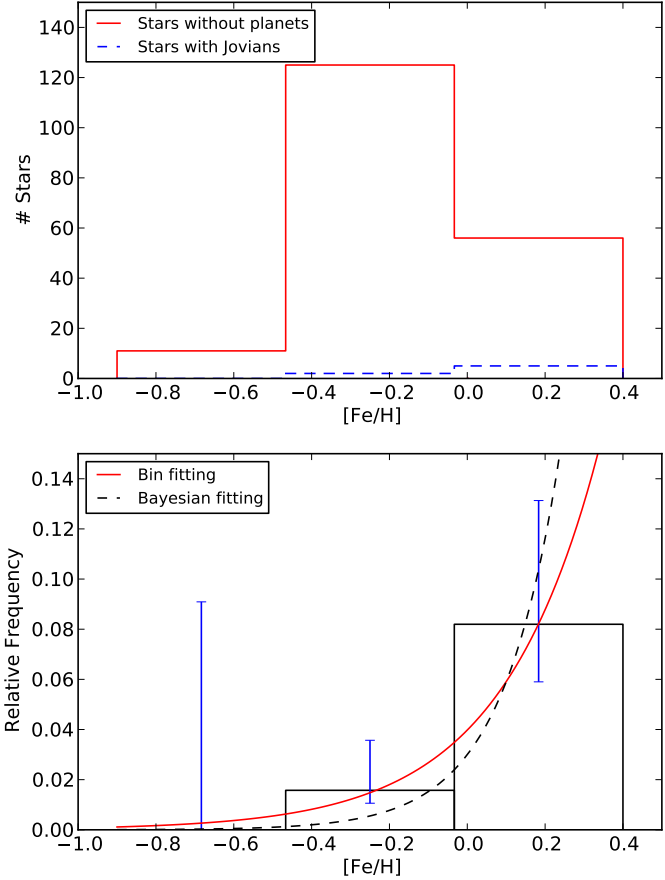}
\end{center}
\caption{Upper panel: Histogram of metallicity of the joined sample with 3 bins for stars without planets(solid red) and stars with Giant planets (dashed blue); Lower panel: Frequency of stars with Giant planets.}
\label{3fehj:joined}
\end{figure}

\begin{figure}[h]
\begin{center}
\includegraphics[scale=0.6]{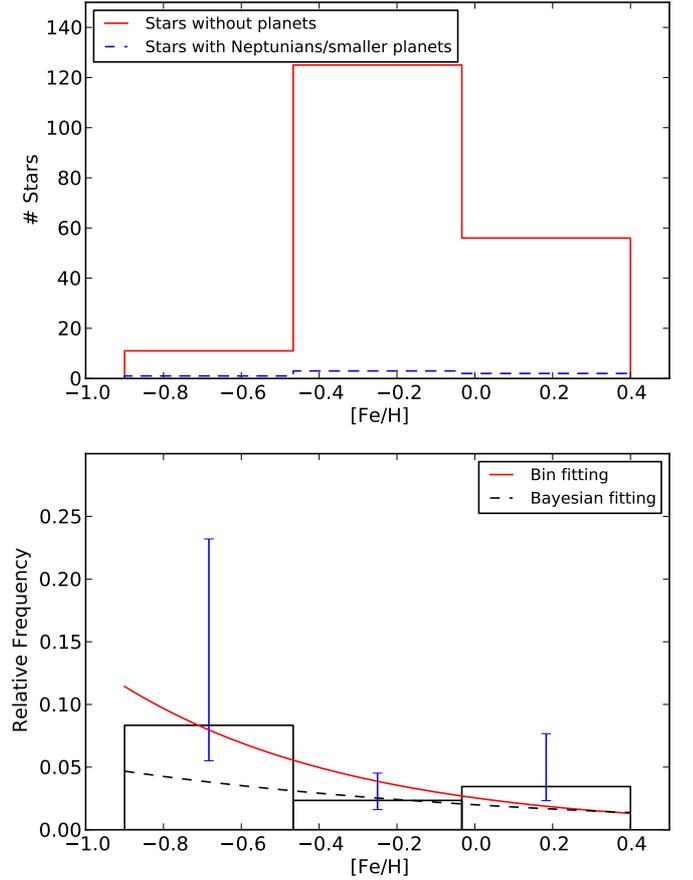}
\end{center}
\caption{Upper panel: Histogram of metallicity of the joined sample with 3 bins for stars without planets(solid red) and stars with Neptunians and smaller planets (dashed blue); Lower panel: Frequency of stars with Neptunians and smaller planets.}
\label{3fehn:joined}
\end{figure}

We now proceed to the frequency analysis of the stars with Jovians and Neptunians/smaller planets. Figures \ref{3fehj:joined} and \ref{3fehn:joined} show, in their upper panel, the histograms of stars with Jovian planets and stars with only Neptunians and smaller planets, respectively, depicted by a dashed blue line. The histogram of the non-host stars of the joined sample are depicted by a solid red line. The lower panels show the frequency of planets of each bin. The solid red and the dashed black lines represent the fit of the binned values and the fit given by a bayesian model (see Sect. \ref{bayes}) respectively. The values of the coefficients for both fits are shown in Table \ref{param:table} and will be discussed together in Sect. \ref{bayes}.

From both figures we can observe that the results are similar to the ones obtained with our sample (see Fig. \ref{3binfehj} and \ref{3binfehn}), but with lower uncertainties. The  correlation of Jovian-hosts and metallicity is now stronger, but the anti-correlation for Neptunians is weaker. The first bin of Fig. \ref{3fehj:joined}, ranging from -0.9 to -0.5 dex has an upper limit of 9.1\%, with no planet detection, while the second and third bins ([-0.5,-0.1] and [-0.1,0.3] dex, resp.) have values of 1.6\% and 8.2\% respectively. Regarding Fig. \ref{3fehn:joined}, we observe the frequencies of 8.3, 2.3, and 3.4\% for the same bins.

%It is interesting to see that with the higher planet-host count we may start to see a build-up of Neptunian and smaller planet host stars at lower metallicities and, as we go towards higher [Fe/H] values we observe the well established [Fe/H]-Giant planet host relation \citep[e.g.][]{Santos-2004b,Fischer-2005}. We must be cautious, however, as the number of planet hosts is still small. The observed frequencies, ranging from 20.0\% to 5.1\% are compatible with the results obtained with our sample.

\subsection{Bayesian approach for the joined sample}
\label{bayes2}

Here we perform the same bayesian inference approach as done in Sect. \ref{bayes} but this time for the joined sample. Table \ref{param:table2} summarizes and compares the results of the Bayesian fitting to the ones obtained with the bin fitting. The columns are the same as in Table \ref{param:table}.

%Column 1 shows the functional forms used and respective parameters, column 2 the uniform prior range, column 3 the most likely value for the fit parameters, along with the $1\sigma$ gaussian uncertainties and column 4 the fit parameters of the least squares bin fitting.

\begin{table}[h]
\caption{Parameters of the two bayesian and fit from binning models for the HARPS+CPS sample.} %\textcolor{red}{To be updated!}}
\label{param:table2}
\resizebox{9cm}{!}{
\begin{tabular}{l | c c c}

\hline
\hline
Parameters  & Uniform  & most likely  & fit from   \\
for Jovian hosts      & Prior       & value & binning \\
\hline
$f_{p1} = C$ & &  & \\
C & (0.01,0.30) & $0.03\pm0.01$ & $0.03\pm0.03$ \\
%\hline
$f_{p2} = C10^{\alpha[Fe/H]}$ & & & \\
C & (0.01,0.30) & $0.03\pm0.02$ & $0.04\pm0.01$ \\
$\alpha$ & (-1.0,4.0) & $2.94\pm1.03$ & $1.72\pm0.18$ \\
\hline
\hline
Parameters  & Uniform  & most likely  & fit from   \\
for Neptunian hosts      & Prior       & value & binning \\
\hline
$f_{p1} = C$ & &  & \\
C & (0.01,0.30) & $0.03\pm0.01$ & $0.04\pm$0.03 \\
%\hline
$f_{p2} = C10^{\alpha[Fe/H]}$ & & & \\
C & (0.01,0.30) & $0.02\pm0.02$ & $0.03\pm0.02$  \\
$\alpha$ & (-4.00,1.00) & $-0.41\pm0.77$ & $-0.72\pm0.46$\\
\hline
\hline\end{tabular}
}
\end{table}

%We note that the $\alpha$ value of the $f_{p} = C10^{\alpha [Fe/H]}$ functional form has large uncertainties. The $\alpha$ value here can easily accommodate both positive or negative values. 

From Table \ref{param:table2} we can see that both the direct bin fitting and the bayesian fitting values are compatible with the ones obtained with the HARPS sample. As we have seen in Sect. \ref{bayes}, the $\alpha$ values are higher than the same values using the bin fitting, translating into a higher Giant-host frequency and a lower Neptunian/smaller planet host frequency. Again, the $\alpha$ values calculated by the Bayesian method have large uncertainties, and the $\alpha$ value, for the Neptunian and smaller planet hosts case, may easily have positive or negative values.

We can now compare the values for giant planets obtained with both fitting methods to previous works. \citet{Valenti-2005}, \citet{Udry-2007}, and \citet{Sousa-2011b} all use a similar power law to the one used in this work for the frequency of giants around FGK dwarfs and obtained $\alpha$ values of 2.0, 2.04, and 2.58 respectively through direct bin fitting. Our $\alpha$ results from the bin fitting ($1.26\pm0.30$ from the HARPS sample and $1.72\pm0.18$ from the joined sample) is lower that those works, which might suggest a less efficient planet-formation process around M dwarfs. However, the $\alpha$ values obtained from the Bayesian fit for the HARPS sample are very similar to the ones obtained for FGK dwarfs: $1.97\pm1.25$, despite the high uncertainty. Regarding the combined sample we obtain a higher value of $2.94\pm1.03$ from the Bayesian fitting, suggesting a more efficient process of planet-formation around M dwarfs. Therefore, our quantification of the $\alpha$ parameter for Giant planets around M dwarfs, taking into account the large uncertainties involved, are compatible with the values found in FGK studies. 

In order to check if the more complex power law functional form is preferred over the constant one, we used a method of Bayesian model comparison, following \citet{Kass-1995}. First, we calculate for both functional forms the total probability of the model conditioned on the data (the evidence) by integrating over the full parameter space. Computationally, in the case of uniformly distributed priors, we can calculate the evidence as  

\begin{equation}
P(d|f) = \frac{\sum{P(d|X)}}{length(X)},
\end{equation}
where the $P(d|X)$ is the likelihood, or the probability of observing the data $d$ given the parameters $X$,  and $length(X)$ is the length of the full parameter space. Then, we calculate the Bayes factor that is just the ratio of the evidence of both functional forms, 

\begin{equation}
B_{f} = \frac{P(d|f_{p2})}{P(d|f_{p1})}.
\end{equation}
According to \citet{Kass-1995} a $B_{f}$ value over 20 gives a \textit{strong} evidence that the model $f_{p2}$ is better at fitting the data than the $f_{p1}$ model. 

For the Jovian hosts case, we obtained a Bayes factor of 2.07 and 66.04 for the HARPS and the joined sample respectively. This means that, in the case of the HARPS sample, the more complex model cannot explain much better the data than the constant model. On the other hand, the combined sample achieves a high Bayes factor, meaning that there is a strong evidence that the more complex model does a better fit than the constant model, supporting the planet-metallicity correlation for Giant planets.

Regarding the Neptunian hosts, we obtain values lower than the unity, which means that the constant model explain the data better than the more complex power model. Therefore, it is impossible at this moment to confirm the hypothetical anti-correlation observed for low [Fe/H] values. Despite this, we must note that our HARPS sample is much more sensitive in probing the Neptunian/Super-earth mass regime than the CPS sample. Therefore the frequency parametrization of the HARPS sample for the Neptunian/Super-earth mass range, and shown in detail in Sect. \ref{ssec:metal}, is preferred over the joined one.

\section{Discussion}
\label{discussion}

In this paper we investigate the metallicity and stellar mass correlations with planets. We use a new method, described in the Appendix, to refine the precision of the metallicities of the HARPS GTO M dwarf sample calculated with the calibration of \citet{Neves-2012}. We use the established calibration of \citet{Delfosse-2000} to calculate the stellar masses of our sample. % based on the photometric calibration of \citet{Neves-2012}, itself a refinement of the \citet{Schlaufman-2010} calibration.} 
%\textcolor{red}{why did you strikeout 'investigate'?}

We confirm the trend of metallicity with the presence of Giant planets in our sample, as shown by previous studies on FGK dwarfs \citep[e.g.][]{Gonzalez-1997, Santos-2004b,Sousa-2011b,Mayor-2011} and M dwarfs \citep{Bonfils-2007, Johnson-2009,Schlaufman-2010,Rojas-Ayala-2012, Terrien-2012}. For Neptunian and smaller planet hosts there is a hint that an  anti-correlation may exist but our current statistic supports a flat relation, in concordance with previous results for FGK dwarfs \citep[e.g.][]{Sousa-2008,Bouchy-2009, Sousa-2011b} and M dwarfs \citep{Rojas-Ayala-2012}. We calculate the difference of the averages and medians between planet and non-planet hosts, and most importantly the frequencies in three different bins, as well as a parametrization to both Jovian and Neptunian hosts. 

We combined the HARPS sample with the California Planet Survey (CPS) late-K and M-type dwarf sample to improve our statistics, increasing the number of stars from 102 to 205 and the number of planet hosts from 8 to 13 (7 Jovian-hosts and 6 Neptunian/smaller planet hosts). The [Fe/H] of the CPS sample was calculated using the photometric calibration of \citet{Neves-2012}. The previous trend for Jovian-hosts is confirmed and reinforced, but the existence of an anti-correlation of Neptunian-hosts with [Fe/H] is inconclusive. The CPS sample is not as sensitive as the HARPS sample regarding the detection of Neptunian and smaller planets. Therefore the HARPS sample is the reference in this work regarding the Neptunian-host-metallicity relation. 

%The giant planet host frequencies now range from 1.8 to 8.2\% while for Neptunians the range is the same as in the HARPS sample. 

Quantitatively, the difference of the averages and the medians between stars with and without planets for Jupiter-type hosts is 0.20 and 0.26 dex for the HARPS sample and 0.20 and 0.19 dex for the joined sample. Regarding the Neptunian and smaller planet hosts, the observed difference of the averages and the medians is, respectively, -0.10 and -0.11 dex for the HARPS sample. %The hint of an anti-correlation of [Fe/H] with planets is at the precision limit of our metallicity technique and is, therefore, inconclusive. The [Fe/H] of the joined sample is not precise enough to measure a Neptunian-host correlation properly. 

Regarding the frequency of Giant hosts, we have no detection in the [-0.9,-0.5] dex bin for both HARPS and the joined sample. For the [-0.5,-0.1] bin we obtained a frequency of 1.9\% and 1.6\%,  and between -0.1 and 0.4 we have a frequency of 5.6\% and 8.2\% for the HARPS and the joined sample respectively. Regarding Neptunian hosts, we obtained, for the same samples and bins, the values of 12.5\% and 8.2\%   for the first bin, 5.3\% and 2.3\% for the second bin, and 2.8\% and 3.4\% for the last [Fe/H] bin. As noted, the frequencies obtained using the joined sample for the Neptunian-hosts are not as precise as in the HARPS sample due to a lower sensitivity of the CPS sample to Neptunian and smaller planets. 

The parametrization of the planet-metallicity relation was based on bin fit and bayesian fit models, following a functional form of the type $f_{p} = C10^{\alpha[Fe/H]}$ used in previous works for FGK dwarfs  \citep[][]{Valenti-2005,Udry-2007,Sousa-2011b}. The results for the parameters $C$ and $\alpha$ using the functional forms calculated by direct bin fitting or by using the Bayesian fitting are compatible with each other. However, we note a high uncertainty on the determination of the $\alpha$ parameter using the Bayesian fitting. Therefore the results for this parameter for Giant planets vary a lot, between $1.26\pm0.30$ and $1.97\pm1.25$, using the bin fitting or the Bayesian fitting respectively, for the HARPS sample, and between $1.72\pm0.18$ to $2.94\pm1.03$ for the combined sample. At the actual statistical level, the $\alpha$ parameter we determine is compatible with the value found for FGK dwarfs in previous studies \citep[e.g.]{Fischer-2005,Udry-2007,Sousa-2011b}. Regarding Neptunian-hosts, we obtain an $\alpha$ value, for the HARPS sample, between $-0.79\pm0.06$ and $-0.57\pm0.71$, using the bin fit or the bayes fit model respectively. This result configures an anti-correlation for Neptunian hosts with [Fe/H], but with an insufficient statistical confidence level. 

We therefore conclude that the power law functional form works best for Giant hosts, and that a constant functional form is preferred, for now, for Neptunian/smaller planet hosts. We also reject the possibility of a correlation for Neptunian-hosts of the same order of magnitude of that for Jupiter-hosts. In fact we suspect that an anti-correlation might exist but we lack the statistics to confirm it.

Regarding stellar mass, we detect a positive trend in planet detections towards higher masses. However, when we take the detection limits into account, we do not find any significant difference. Therefore, the trend of the frequency of planets with the stellar mass is due to a detection bias in our sample, stressing the importance of taking into account the planet detection biases in stellar mass studies.

%Despite that, we need more detections to robustly confirm whether there is a true stellar mass trend or not.

%In this work, we derive the metallicities of a sample of 102 M dwarfs from the HARPS GTO programme. We use a new method, using the high-resolution spectra of HARPS, described in the Appendix, to refine and enhance the precision of metallicities based on the photometric calibration of \citet{Neves-2012}, itself a refinement of the \citet{Schlaufman-2010} calibration. Then, we use this determinations, as well as stellar mass determinations from \citet{Bonfils-2011}, calculated using the mass-luminosity relation of \citet{Delfosse-2000} to search for correlation between the frequency of planets and stellar mass and metallicity. In Sect. \ref{sample}, we describe the sample of M dwarfs and the observations using the HARPS spectrograph. Then, in Sect. \ref{relation}, we investigate the stellar mass/metallcity correlations with the frequency of planets. Afterwards, we calculate the detection limits of the sample, to check for biases in our sample, since the planet detections accumulate in the higher end of stellar mass and lower end of V magnitude. Finally, we discuss our results in Sect. \ref{discussion}.

\begin{acknowledgements}
We would like to thank A. Mortier for useful discussions. We would also like to thank J.A. Johnson and K. Apps for kindly providing the CPS M dwarf sample. We acknowledge the
support by the European Research Council/European Community under the
FP7 through Starting Grant agreement number 239953. The financial support
from the "Programme National de Plan\'etologie'' (PNP) of CNRS/INSU, France, is gratefully acknowledged.
NCS and VN also acknowledges the support from Funda\c{c}\~ao para a Ci\^encia e a
Tecnologia (FCT) through program Ci\^encia\,2007 funded by FCT/MCTES
(Portugal) and POPH/FSE (EC), and in the form of grant reference
PTDC/CTE-AST/098528/2008. VN would also like to acknowledge the
support from the FCT in the form of the fellowship SFRH/BD/60688/2009. 
This research has made use of the SIMBAD database,
operated at CDS, Strasbourg, France, and of the Extrasolar Planet Encyclopaedia at ~ \url{exoplanet.eu}. This publication makes use of data products from the Two Micron All Sky Survey, which is a joint project of the University of Massachusetts and the Infrared Processing and Analysis Center/California Institute of Technology, funded by the National Aeronautics and Space Administration and the National Science Foundation.

\end{acknowledgements}

\bibliographystyle{aa}
\bibliography{mylib.bib}

\begin{thebibliography}{69}
\expandafter\ifx\csname natexlab\endcsname\relax\def\natexlab#1{#1}\fi

\bibitem[{{Alibert} {et~al.}(2011){Alibert}, {Mordasini}, \&
  {Benz}}]{Alibert-2011}
{Alibert}, Y., {Mordasini}, C., \& {Benz}, W. 2011, A\&A, 526, A63

\bibitem[{{Anglada-Escud{\'e}} {et~al.}(2012){Anglada-Escud{\'e}}, {Boss},
  {Weinberger}, {Thompson}, {Butler}, {Vogt}, \&
  {Rivera}}]{Anglada-Escude-2012}
{Anglada-Escud{\'e}}, G., {Boss}, A.~P., {Weinberger}, A.~J., {et~al.} 2012,
  ApJ, 746, 37

\bibitem[{{Baranne} {et~al.}(1996){Baranne}, {Queloz}, {Mayor}, {Adrianzyk},
  {Knispel}, {Kohler}, {Lacroix}, {Meunier}, {Rimbaud}, \&
  {Vin}}]{Baranne-1996}
{Baranne}, A., {Queloz}, D., {Mayor}, M., {et~al.} 1996, A\&AS, 119, 373

\bibitem[{{Benedict} {et~al.}(1999){Benedict}, {McArthur}, {Chappell}, {Nelan},
  {Jefferys}, {van Altena}, {Lee}, {Cornell}, {Shelus}, {Hemenway}, {Franz},
  {Wasserman}, {Duncombe}, {Story}, {Whipple}, \& {Fredrick}}]{Benedict-1999}
{Benedict}, G.~F., {McArthur}, B., {Chappell}, D.~W., {et~al.} 1999, AJ, 118,
  1086

\bibitem[{{Benedict} {et~al.}(2002){Benedict}, {McArthur}, {Forveille},
  {Delfosse}, {Nelan}, {Butler}, {Spiesman}, {Marcy}, {Goldman}, {Perrier},
  {Jefferys}, \& {Mayor}}]{Benedict-2002}
{Benedict}, G.~F., {McArthur}, B.~E., {Forveille}, T., {et~al.} 2002, ApJ, 581,
  L115

\bibitem[{{Bonfils} {et~al.}(2011){Bonfils}, {Delfosse}, {Udry}, {Forveille},
  {Mayor}, {Perrier}, {Bouchy}, {Gillon}, {Lovis}, {Pepe}, {Queloz}, {Santos},
  {S{\'e}gransan}, \& {Bertaux}}]{Bonfils-2011}
{Bonfils}, X., {Delfosse}, X., {Udry}, S., {et~al.} 2011, ArXiv e-prints

\bibitem[{{Bonfils} {et~al.}(2007){Bonfils}, {Mayor}, {Delfosse}, {Forveille},
  {Gillon}, {Perrier}, {Udry}, {Bouchy}, {Lovis}, {Pepe}, {Queloz}, {Santos},
  \& {Bertaux}}]{Bonfils-2007}
{Bonfils}, X., {Mayor}, M., {Delfosse}, X., {et~al.} 2007, A\&A, 474, 293

\bibitem[{{Boss}(1997)}]{Boss-1997}
{Boss}, A.~P. 1997, Science, 276, 1836

\bibitem[{{Boss}(2002)}]{Boss-2002}
{Boss}, A.~P. 2002, ApJ, 567, L149

\bibitem[{{Boss}(2006)}]{Boss-2006a}
{Boss}, A.~P. 2006, ApJ, 643, 501

\bibitem[{{Bouchy} {et~al.}(2009){Bouchy}, {Mayor}, {Lovis}, {Udry}, {Benz},
  {Bertaux}, {Delfosse}, {Mordasini}, {Pepe}, {Queloz}, \&
  {Segransan}}]{Bouchy-2009}
{Bouchy}, F., {Mayor}, M., {Lovis}, C., {et~al.} 2009, A\&A, 496, 527

\bibitem[{{Buchhave} {et~al.}(2012){Buchhave}, {Latham}, {Johansen},
  {Bizzarro}, {Torres}, {Rowe}, {Batalha}, {Borucki}, {Brugamyer}, {Caldwell},
  {Bryson}, {Ciardi}, {Cochran}, {Endl}, {Esquerdo}, {Ford}, {Geary},
  {Gilliland}, {Hansen}, {Isaacson}, {Laird}, {Lucas}, {Marcy}, {Morse},
  {Robertson}, {Shporer}, {Stefanik}, {Still}, \& {Quinn}}]{Buchhave-2012}
{Buchhave}, L.~A., {Latham}, D.~W., {Johansen}, A., {et~al.} 2012, Nature, 486,
  375

\bibitem[{{Burgasser} {et~al.}(2003){Burgasser}, {Kirkpatrick}, {Reid},
  {Brown}, {Miskey}, \& {Gizis}}]{Burgasser-2003}
{Burgasser}, A.~J., {Kirkpatrick}, J.~D., {Reid}, I.~N., {et~al.} 2003, ApJ,
  586, 512

\bibitem[{{Casagrande} {et~al.}(2008){Casagrande}, {Flynn}, \&
  {Bessell}}]{Casagrande-2008}
{Casagrande}, L., {Flynn}, C., \& {Bessell}, M. 2008, MNRAS, 389, 585

\bibitem[{{Correia} {et~al.}(2010){Correia}, {Couetdic}, {Laskar}, {Bonfils},
  {Mayor}, {Bertaux}, {Bouchy}, {Delfosse}, {Forveille}, {Lovis}, {Pepe},
  {Perrier}, {Queloz}, \& {Udry}}]{Correia-2010}
{Correia}, A.~C.~M., {Couetdic}, J., {Laskar}, J., {et~al.} 2010, A\&A, 511,
  A21

\bibitem[{{Delfosse} {et~al.}(2000){Delfosse}, {Forveille}, {S{\'e}gransan},
  {Beuzit}, {Udry}, {Perrier}, \& {Mayor}}]{Delfosse-2000}
{Delfosse}, X., {Forveille}, T., {S{\'e}gransan}, D., {et~al.} 2000, A\&A, 364,
  217

\bibitem[{{Endl} {et~al.}(2006){Endl}, {Cochran}, {K{\"u}rster}, {Paulson},
  {Wittenmyer}, {MacQueen}, \& {Tull}}]{Endl-2006}
{Endl}, M., {Cochran}, W.~D., {K{\"u}rster}, M., {et~al.} 2006, ApJ, 649, 436

\bibitem[{{Fabricius} \& {Makarov}(2000)}]{Fabricius-2000}
{Fabricius}, C. \& {Makarov}, V.~V. 2000, AAPS, 144, 45

\bibitem[{{Fischer} {et~al.}(2005){Fischer}, {Laughlin}, {Butler}, {Marcy},
  {Johnson}, {Henry}, {Valenti}, {Vogt}, {Ammons}, {Robinson}, {Spear},
  {Strader}, {Driscoll}, {Fuller}, {Johnson}, {Manrao}, {McCarthy},
  {Mu{\~n}oz}, {Tah}, {Wright}, {Ida}, {Sato}, {Toyota}, \&
  {Minniti}}]{Fischer-2005b}
{Fischer}, D.~A., {Laughlin}, G., {Butler}, P., {et~al.} 2005, ApJ, 620, 481

\bibitem[{{Fischer} \& {Valenti}(2005)}]{Fischer-2005}
{Fischer}, D.~A. \& {Valenti}, J. 2005, ApJ, 622, 1102

\bibitem[{{Gatewood}(2008)}]{Gatewood-2008}
{Gatewood}, G. 2008, AJ, 136, 452

\bibitem[{{Gatewood} {et~al.}(1998){Gatewood}, {Kiewiet de Jonge}, \&
  {Persinger}}]{Gatewood-1998}
{Gatewood}, G., {Kiewiet de Jonge}, J., \& {Persinger}, T. 1998, AJ, 116, 1501

\bibitem[{{Ghezzi} {et~al.}(2010){Ghezzi}, {Cunha}, {Smith}, {de Ara{\'u}jo},
  {Schuler}, \& {de la Reza}}]{Ghezzi-2010}
{Ghezzi}, L., {Cunha}, K., {Smith}, V.~V., {et~al.} 2010, ApJ, 720, 1290

\bibitem[{{Gonzalez}(1997)}]{Gonzalez-1997}
{Gonzalez}, G. 1997, MNRAS, 285, 403

\bibitem[{{Hawley} {et~al.}(1997){Hawley}, {Gizis}, \& {Reid}}]{Hawley-1997}
{Hawley}, S.~L., {Gizis}, J.~E., \& {Reid}, N.~I. 1997, AJ, 113, 1458

\bibitem[{{Henry} {et~al.}(2006){Henry}, {Jao}, {Subasavage}, {Beaulieu},
  {Ianna}, {Costa}, \& {M{\'e}ndez}}]{Henry-2006}
{Henry}, T.~J., {Jao}, W.-C., {Subasavage}, J.~P., {et~al.} 2006, AJ, 132, 2360

\bibitem[{{Ida} \& {Lin}(2004)}]{Ida-2004b}
{Ida}, S. \& {Lin}, D.~N.~C. 2004, ApJ, 616, 567

\bibitem[{{Ida} \& {Lin}(2005)}]{Ida-2005}
{Ida}, S. \& {Lin}, D.~N.~C. 2005, ApJ, 626, 1045

\bibitem[{{Jahrei{\ss}} \& {Wielen}(1997)}]{Jahreiss-1997}
{Jahrei{\ss}}, H. \& {Wielen}, R. 1997, in ESA Special Publication, Vol. 402,
  Hipparcos - Venice '97, ed. R.~M. {Bonnet}, E.~{H{\o}g}, P.~L. {Bernacca},
  L.~{Emiliani}, A.~{Blaauw}, C.~{Turon}, J.~{Kovalevsky}, L.~{Lindegren},
  H.~{Hassan}, M.~{Bouffard}, B.~{Strim}, D.~{Heger}, M.~A.~C. {Perryman}, \&
  L.~{Woltjer}, 675--680

\bibitem[{{Jao} {et~al.}(2005){Jao}, {Henry}, {Subasavage}, {Brown}, {Ianna},
  {Bartlett}, {Costa}, \& {M{\'e}ndez}}]{Jao-2005}
{Jao}, W.-C., {Henry}, T.~J., {Subasavage}, J.~P., {et~al.} 2005, AJ, 129, 1954

\bibitem[{{Johnson} {et~al.}(2010{\natexlab{a}}){Johnson}, {Aller}, {Howard},
  \& {Crepp}}]{Johnson-2010}
{Johnson}, J.~A., {Aller}, K.~M., {Howard}, A.~W., \& {Crepp}, J.~R.
  2010{\natexlab{a}}, PASP, 122, 905

\bibitem[{{Johnson} \& {Apps}(2009)}]{Johnson-2009}
{Johnson}, J.~A. \& {Apps}, K. 2009, ApJ, 699, 933

\bibitem[{{Johnson} {et~al.}(2007){Johnson}, {Butler}, {Marcy}, {Fischer},
  {Vogt}, {Wright}, \& {Peek}}]{Johnson-2007}
{Johnson}, J.~A., {Butler}, R.~P., {Marcy}, G.~W., {et~al.} 2007, ApJ, 670, 833

\bibitem[{{Johnson} {et~al.}(2010{\natexlab{b}}){Johnson}, {Howard}, {Marcy},
  {Bowler}, {Henry}, {Fischer}, {Apps}, {Isaacson}, \&
  {Wright}}]{Johnson-2010b}
{Johnson}, J.~A., {Howard}, A.~W., {Marcy}, G.~W., {et~al.} 2010{\natexlab{b}},
  PASP, 122, 149

\bibitem[{{Kalas} {et~al.}(2004){Kalas}, {Liu}, \& {Matthews}}]{Kalas-2004}
{Kalas}, P., {Liu}, M.~C., \& {Matthews}, B.~C. 2004, Science, 303, 1990

\bibitem[{{Kass} \& {Raftery}(1995)}]{Kass-1995}
{Kass}, R.~E. \& {Raftery}, A.~E. 1995, Journal of the American Statistical
  Association, 90, 773

\bibitem[{{Kennedy} \& {Kenyon}(2008)}]{Kennedy-2008a}
{Kennedy}, G.~M. \& {Kenyon}, S.~J. 2008, ApJ, 673, 502

\bibitem[{{Kornet} {et~al.}(2006){Kornet}, {Wolf}, \&
  {R{\'o}{\.z}yczka}}]{Kornet-2006}
{Kornet}, K., {Wolf}, S., \& {R{\'o}{\.z}yczka}, M. 2006, A\&A, 458, 661

\bibitem[{{Laws} {et~al.}(2003){Laws}, {Gonzalez}, {Walker}, {Tyagi},
  {Dodsworth}, {Snider}, \& {Suntzeff}}]{Laws-2003}
{Laws}, C., {Gonzalez}, G., {Walker}, K.~M., {et~al.} 2003, AJ, 125, 2664

\bibitem[{{Lovis} \& {Mayor}(2007)}]{Lovis-2007}
{Lovis}, C. \& {Mayor}, M. 2007, A\&A, 472, 657

\bibitem[{{Mayor} {et~al.}(2011){Mayor}, {Marmier}, {Lovis}, {Udry},
  {S{\'e}gransan}, {Pepe}, {Benz}, {Bertaux}, {Bouchy}, {Dumusque}, {Lo Curto},
  {Mordasini}, {Queloz}, \& {Santos}}]{Mayor-2011}
{Mayor}, M., {Marmier}, M., {Lovis}, C., {et~al.} 2011, ArXiv e-prints

\bibitem[{{Mayor} {et~al.}(2003){Mayor}, {Pepe}, {Queloz}, {Bouchy},
  {Rupprecht}, {Lo Curto}, {Avila}, {Benz}, {Bertaux}, {Bonfils}, {dall},
  {Dekker}, {Delabre}, {Eckert}, {Fleury}, {Gilliotte}, {Gojak}, {Guzman},
  {Kohler}, {Lizon}, {Longinotti}, {Lovis}, {Megevand}, {Pasquini}, {Reyes},
  {Sivan}, {Sosnowska}, {Soto}, {Udry}, {van Kesteren}, {Weber}, \&
  {Weilenmann}}]{Mayor-2003b}
{Mayor}, M., {Pepe}, F., {Queloz}, D., {et~al.} 2003, The Messenger, 114, 20

\bibitem[{{McCarthy} \& {Zuckerman}(2004)}]{McCarthy-2004}
{McCarthy}, C. \& {Zuckerman}, B. 2004, AJ, 127, 2871

\bibitem[{{Melo} {et~al.}(2007){Melo}, {Santos}, {Gieren}, {Pietrzynski},
  {Ruiz}, {Sousa}, {Bouchy}, {Lovis}, {Mayor}, {Pepe}, {Queloz}, {Da Silva}, \&
  {Udry}}]{Melo-2007}
{Melo}, C., {Santos}, N.~C., {Gieren}, W., {et~al.} 2007, A\&A, 467, 721

\bibitem[{{Mordasini} {et~al.}(2009){Mordasini}, {Alibert}, \&
  {Benz}}]{Mordasini-2009}
{Mordasini}, C., {Alibert}, Y., \& {Benz}, W. 2009, A\&A, 501, 1139

\bibitem[{{Mordasini} {et~al.}(2012){Mordasini}, {Alibert}, {Benz}, {Klahr}, \&
  {Henning}}]{Mordasini-2012}
{Mordasini}, C., {Alibert}, Y., {Benz}, W., {Klahr}, H., \& {Henning}, T. 2012,
  ArXiv e-prints

\bibitem[{{Murray} {et~al.}(2002){Murray}, {Paskowitz}, \&
  {Holman}}]{Murray-2002}
{Murray}, N., {Paskowitz}, M., \& {Holman}, M. 2002, ApJ, 565, 608

\bibitem[{{Neves} {et~al.}(2012){Neves}, {Bonfils}, {Santos}, {Delfosse},
  {Forveille}, {Allard}, {Nat{\'a}rio}, {Fernandes}, \& {Udry}}]{Neves-2012}
{Neves}, V., {Bonfils}, X., {Santos}, N.~C., {et~al.} 2012, a\&a, 538, A25

\bibitem[{{Pepe} {et~al.}(2004){Pepe}, {Mayor}, {Queloz}, {Benz}, {Bonfils},
  {Bouchy}, {Curto}, {Lovis}, {M{\' e}gevand}, {Moutou}, {Naef}, {Rupprecht},
  {Santos}, {Sivan}, {Sosnowska}, \& {Udry}}]{Pepe-2004}
{Pepe}, F., {Mayor}, M., {Queloz}, D., {et~al.} 2004, A\&A, 423, 385

\bibitem[{{Pollack} {et~al.}(1996){Pollack}, {Hubickyj}, {Bodenheimer},
  {Lissauer}, {Podolak}, \& {Greenzweig}}]{Pollack-1996}
{Pollack}, J., {Hubickyj}, O., {Bodenheimer}, P., {et~al.} 1996, Icarus, 124,
  62

\bibitem[{{Rauscher} \& {Marcy}(2006)}]{Rauscher-2006}
{Rauscher}, E. \& {Marcy}, G.~W. 2006, PASP, 118, 617

\bibitem[{{Reid} {et~al.}(1995){Reid}, {Hawley}, \& {Gizis}}]{Reid-1995}
{Reid}, I.~N., {Hawley}, S.~L., \& {Gizis}, J.~E. 1995, AJ, 110, 1838

\bibitem[{{Rivera} {et~al.}(2010){Rivera}, {Laughlin}, {Butler}, {Vogt},
  {Haghighipour}, \& {Meschiari}}]{Rivera-2010}
{Rivera}, E.~J., {Laughlin}, G., {Butler}, R.~P., {et~al.} 2010, ApJ, 719, 890

\bibitem[{{Rojas-Ayala} {et~al.}(2010){Rojas-Ayala}, {Covey}, {Muirhead}, \&
  {Lloyd}}]{Rojas-Ayala-2010}
{Rojas-Ayala}, B., {Covey}, K.~R., {Muirhead}, P.~S., \& {Lloyd}, J.~P. 2010,
  ApJ, 720, L113

\bibitem[{{Rojas-Ayala} {et~al.}(2012){Rojas-Ayala}, {Covey}, {Muirhead}, \&
  {Lloyd}}]{Rojas-Ayala-2012}
{Rojas-Ayala}, B., {Covey}, K.~R., {Muirhead}, P.~S., \& {Lloyd}, J.~P. 2012,
  \apj, 748, 93

\bibitem[{{Santos} {et~al.}(2004){Santos}, {Israelian}, \&
  {Mayor}}]{Santos-2004b}
{Santos}, N.~C., {Israelian}, G., \& {Mayor}, M. 2004, A\&A, 415, 1153

\bibitem[{{Schlaufman} \& {Laughlin}(2010)}]{Schlaufman-2010}
{Schlaufman}, K.~C. \& {Laughlin}, G. 2010, A\&A, 519, A105+

\bibitem[{{Skrutskie} {et~al.}(2006){Skrutskie}, {Cutri}, {Stiening},
  {Weinberg}, {Schneider}, {Carpenter}, {Beichman}, {Capps}, {Chester},
  {Elias}, {Huchra}, {Liebert}, {Lonsdale}, {Monet}, {Price}, {Seitzer},
  {Jarrett}, {Kirkpatrick}, {Gizis}, {Howard}, {Evans}, {Fowler}, {Fullmer},
  {Hurt}, {Light}, {Kopan}, {Marsh}, {McCallon}, {Tam}, {Van Dyk}, \&
  {Wheelock}}]{Skrutskie-2006}
{Skrutskie}, M.~F., {Cutri}, R.~M., {Stiening}, R., {et~al.} 2006, AJ, 131,
  1163

\bibitem[{{S{\"o}derhjelm}(1999)}]{Soderhjelm-1999}
{S{\"o}derhjelm}, S. 1999, \aap, 341, 121

\bibitem[{{Sousa} {et~al.}(2011){Sousa}, {Santos}, {Israelian}, {Mayor}, \&
  {Udry}}]{Sousa-2011b}
{Sousa}, S.~G., {Santos}, N.~C., {Israelian}, G., {Mayor}, M., \& {Udry}, S.
  2011, \aap, 533, A141

\bibitem[{{Sousa} {et~al.}(2008){Sousa}, {Santos}, {Mayor}, {Udry},
  {Casagrande}, {Israelian}, {Pepe}, {Queloz}, \& {Monteiro}}]{Sousa-2008}
{Sousa}, S.~G., {Santos}, N.~C., {Mayor}, M., {et~al.} 2008, A\&A, 487, 373

\bibitem[{{Sozzetti} {et~al.}(2009){Sozzetti}, {Torres}, {Latham}, {Stefanik},
  {Korzennik}, {Boss}, {Carney}, \& {Laird}}]{Sozzetti-2009}
{Sozzetti}, A., {Torres}, G., {Latham}, D.~W., {et~al.} 2009, ApJ, 697, 544

\bibitem[{{Terrien} {et~al.}(2012){Terrien}, {Mahadevan}, {Bender},
  {Deshpande}, {Ramsey}, \& {Bochanski}}]{Terrien-2012}
{Terrien}, R.~C., {Mahadevan}, S., {Bender}, C.~F., {et~al.} 2012, \apjl, 747,
  L38

\bibitem[{{Thommes} {et~al.}(2008){Thommes}, {Matsumura}, \&
  {Rasio}}]{Thommes-2008}
{Thommes}, E.~W., {Matsumura}, S., \& {Rasio}, F.~A. 2008, Science, 321, 814

\bibitem[{{Udry} \& {Santos}(2007)}]{Udry-2007}
{Udry}, S. \& {Santos}, N. 2007, ARAA, 45, 397

\bibitem[{{Valenti} \& {Fischer}(2005)}]{Valenti-2005}
{Valenti}, J.~A. \& {Fischer}, D.~A. 2005, VizieR Online Data Catalog, 215,
  90141

\bibitem[{{van Altena} {et~al.}(1995){van Altena}, {Lee}, \&
  {Hoffleit}}]{van_Altena-1995}
{van Altena}, W.~F., {Lee}, J.~T., \& {Hoffleit}, E.~D. 1995, {The general
  catalogue of trigonometric [stellar] parallaxes}, ed. {van Altena, W.~F.,
  Lee, J.~T., \& Hoffleit, E.~D.}

\bibitem[{{van Leeuwen}(2007)}]{van_Leeuwen-2007}
{van Leeuwen}, F. 2007, A\&A, 474, 653

\bibitem[{{Vorobyov} \& {Basu}(2008)}]{Vorobyov-2008}
{Vorobyov}, E.~I. \& {Basu}, S. 2008, ApJL, 676, L139

\end{thebibliography}

\onecolumn 

\longtab{2}{
%\begin{longtable}[b]
%\begin{center}
%\resizebox{9cm}{!}{
\begin{longtable}{ l r r r c r r r r r}

\caption{\label{full_table} HARPS M dwarf sample ample. Sorted by right ascension.} \\

%\toplabel{blabla}	
%\begin{firsthead}
\hline
\hline
Star & $\alpha$ (2000) & $\delta$ (2000) & $\pi$ & $\pi$ src & Stype & V & $K_{S}$ & $M_{\star}$ & [Fe/H] \\
 &  & & [mas] & & & [mag] & [mag] & $[M_{\odot}]$ & [dex]\\
\hline
\endfirsthead
\caption{continued.}\\
\hline
\hline
Star & $\alpha$ (2000) & $\delta$ (2000) & $\pi$ & $\pi$ src & Stype & V & $K_{S}$ & $M_{\star}$ & [Fe/H] \\
 &  & & [mas] & & & [mag] & [mag] & $[M_{\odot}]$ & [dex]\\
\hline
\endhead
\hline

\endfoot
Gl1 & 00:05:25 & -37:21:23 & $230.4\pm 0.9$ & H & M3V &  8.6 & $4.501\pm0.030$ & $0.39\pm0.03$ & -0.45 \\
GJ1002 & 00:06:44 & -07:32:23 & $213.0\pm 3.6$ & H & M5.5V & 13.8 & $7.439\pm0.021$ & $0.11\pm0.01$ & -0.19 \\
Gl12 & 00:15:49 & +13:33:17 & $88.8\pm 3.5$ & H & M3 & 12.6 & $7.807\pm0.020$ & $0.22\pm0.02$ & -0.34 \\
LHS1134 & 00:43:26 & -41:17:36 & $101.0\pm16.0$ & R & M3 & 13.1 & $7.710\pm0.016$ & $0.20\pm0.01$ & -0.10 \\
Gl54.1 & 01:12:31 & -17:00:00 & $271.0\pm 8.4$ & H & M4.5V & 12.0 & $6.420\pm0.017$ & $0.13\pm0.01$ & -0.40 \\
L707-74 & 01:23:18 & -12:56:23 & $97.8\pm13.5$ & Y & M & 13.0 & $8.350\pm0.021$ & $0.15\pm0.02$ & -0.35 \\
Gl87 & 02:12:21 & +03:34:30 & $96.0\pm 1.7$ & H & M1.5 & 10.1 & $6.077\pm0.020$ & $0.45\pm0.03$ & -0.31 \\
Gl105B & 02:36:16 & +06:52:12 & $139.3\pm 0.5$ & H & M3.5V & 11.7 & $6.574\pm0.020$ & $0.25\pm0.02$ & -0.02 \\
CD-44-836A & 02:45:11 & -43:44:30 & $113.9\pm38.7$ & C & M4 & 12.3 & $7.270\pm0.024$ & $0.22\pm0.02$ & -0.08 \\
LHS1481 & 02:58:10 & -12:53:06 & $95.5\pm10.9$ & H & M2.5 & 12.7 & $8.199\pm0.026$ & $0.17\pm0.02$ & -0.72 \\
LP771-95A & 03:01:51 & -16:35:36 & $146.4\pm 2.9$ & H06 & M3 & 11.5 & $6.285\pm0.020$ & $0.24\pm0.02$ & -0.34 \\
LHS1513 & 03:11:36 & -38:47:17 & $130.0\pm20.0$ & R & M3.5 & 11.5 & $9.016\pm0.022$ & $0.09\pm0.02$ & -0.11 \\
GJ1057 & 03:13:23 & +04:46:30 & $117.1\pm 3.5$ & H & M5 & 13.9 & $7.833\pm0.024$ & $0.16\pm0.01$ & 0.10 \\
Gl145 & 03:32:56 & -44:42:06 & $93.1\pm 1.9$ & H & M2.5 & 11.5 & $6.907\pm0.016$ & $0.32\pm0.02$ & -0.28 \\
GJ1061 & 03:36:00 & -44:30:48 & $271.9\pm 1.3$ & H & M5.5V & 13.1 & $6.610\pm0.021$ & $0.12\pm0.01$ & -0.08 \\
GJ1065 & 03:50:44 & -06:05:42 & $105.4\pm 3.2$ & H & M4V & 12.8 & $7.751\pm0.020$ & $0.19\pm0.02$ & -0.22 \\
GJ1068 & 04:10:28 & -53:36:06 & $143.4\pm 1.9$ & H & M4.5 & 13.6 & $7.900\pm0.021$ & $0.13\pm0.01$ & -0.30 \\
Gl166C & 04:15:22 & -07:39:23 & $200.6\pm 0.2$ & H & M4.5V & 11.2 & $5.962\pm0.026$ & $0.23\pm0.02$ & 0.08 \\
Gl176 & 04:42:56 & +18:57:29 & $106.2\pm 2.5$ & H & M2.5 & 10.0 & $4.310\pm0.034$ & $0.50\pm0.03$ & -0.01 \\
LHS1723 & 05:01:57 & -06:56:47 & $187.9\pm 1.3$ & H & M3.5V & 12.2 & $6.736\pm0.024$ & $0.17\pm0.01$ & -0.25 \\
LHS1731 & 05:03:20 & -17:22:23 & $108.6\pm 2.7$ & H & M3.0V & 11.7 & $6.936\pm0.021$ & $0.27\pm0.02$ & -0.26 \\
Gl191 & 05:11:40 & -45:01:06 & $255.3\pm 0.9$ & H & M1 pV &  8.8 & $5.049\pm0.021$ & $0.27\pm0.03$ & -0.88 \\
Gl203 & 05:28:00 & +09:38:36 & $113.5\pm 5.0$ & H & M3.5V & 12.4 & $7.542\pm0.017$ & $0.19\pm0.02$ & -0.25 \\
Gl205 & 05:31:27 & -03:40:42 & $176.8\pm 1.2$ & H & M1.5V &  8.0 & $4.039\pm0.260$ & $0.60\pm0.07$ & 0.22 \\
Gl213 & 05:42:09 & +12:29:23 & $171.6\pm 4.0$ & H & M4V & 11.5 & $6.389\pm0.016$ & $0.22\pm0.02$ & -0.11 \\
Gl229 & 06:10:34 & -21:51:53 & $173.8\pm 1.0$ & H & M1V &  8.2 & $4.166\pm0.232$ & $0.58\pm0.06$ & -0.01 \\
HIP31293 & 06:33:43 & -75:37:47 & $110.9\pm 2.2$ & H & M3V & 10.5 & $5.862\pm0.024$ & $0.43\pm0.03$ & -0.04 \\
HIP31292 & 06:33:47 & -75:37:30 & $114.5\pm 3.2$ & H & M3/4V & 11.4 & $6.558\pm0.021$ & $0.31\pm0.02$ & -0.10 \\
G108-21 & 06:42:11 & +03:34:53 & $103.1\pm 8.5$ & H & M3.5 & 12.1 & $7.334\pm0.031$ & $0.23\pm0.02$ & -0.01 \\
Gl250B & 06:52:18 & -05:11:24 & $114.8\pm 0.4$ & H & M2.5V & 10.1 & $5.723\pm0.036$ & $0.45\pm0.03$ & -0.10 \\
Gl273 & 07:27:24 & +05:13:30 & $263.0\pm 1.4$ & H & M3.5V &  9.8 & $4.857\pm0.023$ & $0.29\pm0.02$ & -0.01 \\
LHS1935 & 07:38:41 & -21:13:30 & $94.3\pm 3.3$ & H & M3 & 11.7 & $7.063\pm0.023$ & $0.29\pm0.02$ & -0.24 \\
Gl285 & 07:44:40 & +03:33:06 & $167.9\pm 2.3$ & H & M4V & 11.2 & $5.698\pm0.017$ & $0.31\pm0.02$ & 0.18 \\
Gl299 & 08:11:57 & +08:46:23 & $146.3\pm 3.1$ & H & M4V & 12.8 & $7.660\pm0.026$ & $0.14\pm0.01$ & -0.50 \\
Gl300 & 08:12:41 & -21:33:12 & $125.8\pm 1.0$ & H & M3.5V & 12.1 & $6.705\pm0.027$ & $0.26\pm0.02$ & 0.14 \\
GJ2066 & 08:16:08 & +01:18:11 & $109.6\pm 1.5$ & H & M2 & 10.1 & $5.766\pm0.024$ & $0.46\pm0.03$ & -0.18 \\
GJ1123 & 09:17:05 & -77:49:17 & $110.9\pm 2.0$ & H & M4.5V & 13.1 & $7.448\pm0.021$ & $0.21\pm0.01$ & 0.20 \\
Gl341 & 09:21:38 & -60:16:53 & $95.6\pm 0.9$ & H & M0V &  9.5 & $5.587\pm0.021$ & $0.55\pm0.03$ & -0.13 \\
GJ1125 & 09:30:44 & +00:19:18 & $103.5\pm 3.9$ & H & M3.0V & 11.7 & $6.871\pm0.024$ & $0.29\pm0.02$ & -0.30 \\
Gl357 & 09:36:02 & -21:39:42 & $110.8\pm 1.9$ & H & M3V & 10.9 & $6.475\pm0.017$ & $0.33\pm0.03$ & -0.34 \\
Gl358 & 09:39:47 & -41:04:00 & $105.6\pm 1.6$ & H & M3.0V & 10.8 & $6.056\pm0.023$ & $0.42\pm0.03$ & -0.01 \\
Gl367 & 09:44:30 & -45:46:36 & $101.3\pm 3.2$ & H & M1 & 10.1 & $5.780\pm0.020$ & $0.49\pm0.03$ & -0.07 \\
GJ1129 & 09:44:48 & -18:12:48 & $90.9\pm 3.8$ & H & M3.5V & 12.5 & $7.257\pm0.020$ & $0.28\pm0.02$ & 0.07 \\
Gl382 & 10:12:17 & -03:44:47 & $127.1\pm 1.9$ & H & M2V &  9.3 & $5.015\pm0.020$ & $0.54\pm0.03$ & 0.04 \\
Gl388 & 10:19:36 & +19:52:12 & $204.6\pm 2.8$ & H & M4.5 &  9.4 & $4.593\pm0.017$ & $0.42\pm0.03$ & 0.07 \\
Gl393 & 10:28:55 & +00:50:23 & $141.5\pm 2.2$ & H & M2V &  9.7 & $5.311\pm0.023$ & $0.44\pm0.03$ & -0.22 \\
LHS288 & 10:44:32 & -61:11:35 & $209.7\pm 2.7$ & H & M5.5 & 13.9 & $7.728\pm0.027$ & $0.10\pm0.01$ & -0.60 \\
Gl402 & 10:50:52 & +06:48:30 & $147.9\pm 3.5$ & H & M4V & 11.7 & $6.371\pm0.016$ & $0.26\pm0.02$ & 0.06 \\
Gl406 & 10:56:29 & +07:00:54 & $419.1\pm 2.1$ & H & M6V & 13.4 & $6.084\pm0.017$ & $0.10\pm0.00$ & 0.18 \\
Gl413.1 & 11:09:31 & -24:36:00 & $93.0\pm 1.7$ & H & M2 & 10.4 & $6.097\pm0.023$ & $0.46\pm0.03$ & -0.12 \\
Gl433 & 11:35:27 & -32:32:23 & $112.6\pm 1.4$ & H & M2.0V &  9.8 & $5.623\pm0.021$ & $0.47\pm0.03$ & -0.17 \\
Gl438 & 11:43:20 & -51:50:23 & $119.0\pm10.2$ & R & M0 & 10.4 & $6.320\pm0.021$ & $0.33\pm0.03$ & -0.39 \\
Gl447 & 11:47:44 & +00:48:16 & $299.6\pm 2.2$ & H & M4 & 11.1 & $5.654\pm0.024$ & $0.17\pm0.01$ & -0.18 \\
Gl465 & 12:24:53 & -18:14:30 & $113.0\pm 2.5$ & H & M3V & 11.3 & $6.950\pm0.021$ & $0.26\pm0.02$ & -0.66 \\
Gl479 & 12:37:53 & -52:00:06 & $103.2\pm 2.3$ & H & M3V & 10.7 & $6.020\pm0.021$ & $0.43\pm0.03$ & 0.02 \\
LHS337 & 12:38:50 & -38:22:53 & $156.8\pm 2.0$ & H & M4.5V & 12.7 & $7.386\pm0.021$ & $0.15\pm0.01$ & -0.25 \\
Gl480.1 & 12:40:46 & -43:34:00 & $128.5\pm 3.9$ & H & M3.0V & 12.2 & $7.413\pm0.021$ & $0.18\pm0.02$ & -0.48 \\
Gl486 & 12:47:57 & +09:45:12 & $119.5\pm 2.7$ & H & M3.5 & 11.4 & $6.362\pm0.018$ & $0.32\pm0.02$ & 0.06 \\
Gl514 & 13:30:00 & +10:22:36 & $130.6\pm 1.1$ & H & M1V &  9.1 & $5.036\pm0.027$ & $0.53\pm0.03$ & -0.16 \\
Gl526 & 13:45:44 & +14:53:30 & $185.5\pm 1.1$ & H & M1.5V &  8.5 & $4.415\pm0.017$ & $0.50\pm0.03$ & -0.20 \\
Gl536 & 14:01:03 & -02:39:18 & $98.3\pm 1.6$ & H & M1 &  9.7 & $5.683\pm0.020$ & $0.52\pm0.03$ & -0.12 \\
Gl551 & 14:29:43 & -62:40:47 & $771.6\pm 2.6$ & H & M5.5 & 11.1 & $4.310\pm0.030$ & $0.12\pm0.01$ & -0.00 \\
Gl555 & 14:34:17 & -12:31:06 & $165.0\pm 3.3$ & H & M3.5V & 11.3 & $5.939\pm0.034$ & $0.28\pm0.02$ & 0.17 \\
Gl569A & 14:54:29 & +16:06:04 & $101.9\pm 1.7$ & H & M2.5 & 10.2 & $5.770\pm0.018$ & $0.49\pm0.03$ & -0.08 \\
Gl581 & 15:19:26 & -07:43:17 & $160.9\pm 2.6$ & H & M2.5V & 10.6 & $5.837\pm0.023$ & $0.30\pm0.02$ & -0.21 \\
Gl588 & 15:32:13 & -41:16:36 & $168.7\pm 1.3$ & H & M2.5V &  9.3 & $4.759\pm0.024$ & $0.47\pm0.03$ & 0.07 \\
Gl618A & 16:20:04 & -37:31:41 & $119.8\pm 2.5$ & H & M3V & 10.6 & $5.950\pm0.021$ & $0.39\pm0.03$ & -0.08 \\
Gl628 & 16:30:18 & -12:39:47 & $233.0\pm 1.6$ & H & M3V & 10.1 & $5.075\pm0.024$ & $0.30\pm0.02$ & -0.02 \\
Gl643 & 16:55:25 & -08:19:23 & $148.9\pm 4.0$ & H & M3.5V & 11.8 & $6.724\pm0.017$ & $0.21\pm0.02$ & -0.28 \\
Gl667C & 17:18:58 & -34:59:42 & $146.3\pm 9.0$ & H & M2V & 10.2 & $6.036\pm0.020$ & $0.30\pm0.03$ & -0.53 \\
Gl674 & 17:28:40 & -46:53:42 & $220.2\pm 1.4$ & H & M3V &  9.4 & $4.855\pm0.018$ & $0.35\pm0.03$ & -0.25 \\
Gl678.1A & 17:30:22 & +05:32:53 & $100.2\pm 1.1$ & H & M1V &  9.3 & $5.422\pm0.029$ & $0.57\pm0.03$ & -0.11 \\
Gl680 & 17:35:13 & -48:40:53 & $102.8\pm 2.8$ & H & M1.5 & 10.2 & $5.829\pm0.021$ & $0.47\pm0.03$ & -0.22 \\
Gl682 & 17:37:03 & -44:19:11 & $196.9\pm 2.1$ & H & M4.5V & 11.0 & $5.606\pm0.020$ & $0.27\pm0.02$ & 0.11 \\
Gl686 & 17:37:53 & +18:35:30 & $123.0\pm 1.6$ & H & M1 &  9.6 & $5.572\pm0.020$ & $0.45\pm0.03$ & -0.37 \\
Gl693 & 17:46:35 & -57:19:11 & $171.5\pm 2.3$ & H & M3.5V & 10.8 & $6.016\pm0.017$ & $0.26\pm0.02$ & -0.30 \\
Gl699 & 17:57:49 & +04:41:36 & $549.0\pm 1.6$ & H & M4V &  9.6 & $4.524\pm0.020$ & $0.16\pm0.01$ & -0.52 \\
Gl701 & 18:05:07 & -03:01:53 & $128.9\pm 1.4$ & H & M0V &  9.4 & $5.306\pm0.021$ & $0.48\pm0.03$ & -0.27 \\
GJ1224 & 18:07:33 & -15:57:47 & $132.6\pm 3.7$ & H & M4.5V & 13.6 & $7.827\pm0.027$ & $0.14\pm0.01$ & -0.10 \\
G141-29 & 18:42:44 & +13:54:17 & $93.3\pm11.5$ & H & M4 & 12.8 & $7.551\pm0.021$ & $0.23\pm0.02$ & 0.09 \\
Gl729 & 18:49:49 & -23:50:12 & $336.7\pm 2.0$ & H & M3.5V & 10.5 & $5.370\pm0.016$ & $0.17\pm0.01$ & -0.10 \\
GJ1232 & 19:09:51 & +17:40:07 & $93.6\pm 2.8$ & H & M4.5 & 13.6 & $7.902\pm0.020$ & $0.20\pm0.01$ & 0.14 \\
Gl752A & 19:16:55 & +05:10:05 & $170.4\pm 1.0$ & H & M3V &  9.1 & $4.673\pm0.020$ & $0.48\pm0.03$ & 0.06 \\
Gl754 & 19:20:48 & -45:33:30 & $169.2\pm 1.6$ & H & M4.5 & 12.2 & $6.845\pm0.026$ & $0.18\pm0.01$ & -0.17 \\
GJ1236 & 19:22:03 & +07:02:36 & $92.9\pm 2.5$ & H & M3 & 12.4 & $7.688\pm0.020$ & $0.22\pm0.02$ & -0.42 \\
GJ1256 & 20:40:34 & +15:29:57 & $102.0\pm 2.2$ & H & M4.5 & 13.4 & $7.749\pm0.031$ & $0.19\pm0.01$ & 0.10 \\
Gl803$^{\dag}$ & 20:45:10 & -31:20:30 & $100.9\pm 1.1$ & H & M0V e &  8.8 & $4.529\pm0.020$ & $0.75\pm0.03$ & 0.32 \\
LHS3583 & 20:46:37 & -81:43:12 & $77.1\pm21.2$ & C & M2.5 & 11.5 & $6.826\pm0.034$ & $0.40\pm0.03$ & -0.18 \\
LP816-60 & 20:52:33 & -16:58:30 & $175.0\pm 3.4$ & H & M & 11.4 & $6.199\pm0.021$ & $0.23\pm0.02$ & -0.06 \\
Gl832 & 21:33:34 & -49:00:36 & $201.9\pm 1.0$ & H & M1V &  8.7 & $4.473\pm0.050$ & $0.45\pm0.03$ & -0.19 \\
Gl846 & 22:02:10 & +01:24:00 & $97.6\pm 1.5$ & H & M0.5V &  9.2 & $5.322\pm0.023$ & $0.60\pm0.03$ & 0.06 \\
LHS3746 & 22:02:29 & -37:04:54 & $134.3\pm 1.3$ & H & M3.5 & 11.8 & $6.718\pm0.020$ & $0.24\pm0.02$ & -0.15 \\
Gl849 & 22:09:40 & -04:38:30 & $109.9\pm 2.1$ & H & M3V & 10.4 & $5.594\pm0.017$ & $0.49\pm0.03$ & 0.24 \\
GJ1265 & 22:13:42 & -17:41:12 & $96.0\pm 3.9$ & H & M4.5 & 13.6 & $8.115\pm0.018$ & $0.17\pm0.01$ & -0.09 \\
LHS3799 & 22:23:07 & -17:36:23 & $134.4\pm 4.9$ & H & M4.5V & 13.3 & $7.319\pm0.018$ & $0.18\pm0.01$ & 0.18 \\
Gl876 & 22:53:17 & -14:15:48 & $213.3\pm 2.1$ & H & M3.5V & 10.2 & $5.010\pm0.021$ & $0.34\pm0.02$ & 0.15 \\
Gl877 & 22:55:46 & -75:27:36 & $116.1\pm 1.2$ & H & M2.5 & 10.4 & $5.811\pm0.021$ & $0.43\pm0.03$ & -0.01 \\
Gl880 & 22:56:35 & +16:33:12 & $146.1\pm 1.0$ & H & M1.5V &  8.7 & $4.523\pm0.016$ & $0.58\pm0.03$ & 0.07 \\
Gl887 & 23:05:52 & -35:51:12 & $303.9\pm 0.9$ & H & M2V &  7.3 & $3.465\pm0.200$ & $0.47\pm0.05$ & -0.24 \\
LHS543 & 23:21:37 & +17:17:25 & $91.0\pm 2.9$ & H & M4 & 11.7 & $6.507\pm0.016$ & $0.40\pm0.02$ & 0.25 \\
Gl908 & 23:49:13 & +02:24:06 & $167.3\pm 1.2$ & H & M1V &  9.0 & $5.043\pm0.020$ & $0.42\pm0.03$ & -0.44 \\
LTT9759 & 23:53:50 & -75:37:53 & $100.1\pm 1.1$ & H & M & 10.0 & $5.549\pm0.027$ & $0.54\pm0.03$ & 0.21 \\
%\hline
%\hline
\end{longtable}
\raggedright{
$\pi$ src: (H) revised Hipparcos catalog \citep{van_Leeuwen-2007}; (R95) \citep{Reid-1995}; (Y) \citep{van_Altena-1995}; (H06) \citep{Henry-2006}; (C) CNS4 catalog (Jahreiss, private comm.) \\
$^{\dag}$ Gl803 is a young ($\sim$20 Myr) M dwarf with a circumstellar disk \citep{Kalas-2004}. The equation to determine its mass may not be adequate for this age. \\
}

%\end{center}
}

\longtab{8}{
\begin{longtable}{l r r r c r r r r r r}
\caption{\label{full_table_CPS} California Planet Survey (CPS) sample. Sorted by right ascension.} \\
\hline
\hline
Star & $\alpha$ (2000) & $\delta$ (2000) &$\pi$ & $\pi_{src}$ & Stype & V & $K_{S}$ & $M_{\star}$ & [Fe/H]$_{JA09}$ & [Fe/H]$_{N12}$ \\
 & & & [mas] & & & [mag] & [mag] & $[M_{\odot}]$ & [dex] & [dex] \\
\hline
\endfirsthead
\caption{continued.} \\
\hline
\hline
Star & $\alpha$ (2000) & $\delta$ (2000) &$\pi$ & $\pi_{src}$ & Stype & V & $K_{S}$ & $M_{\star}$ & [Fe/H]$_{JA09}$ & [Fe/H]$_{N12}$ \\
 & & & [mas] & & [mag] & [mag] & $[M_{\odot}]$ & [dex]\\
\hline
\endhead
GJ2 & 00:05:10 & 45:47:11 & $88.9\pm 1.4$ & H & M1 &  9.9 & $5.853\pm0.018$ & $0.53\pm0.03$ & 0.06 & -0.09 \\
GJ1 & 00:05:24 & -37:21:26 & $230.4\pm 0.9$ & H & M1.5 &  8.6 & $4.523\pm0.017$ & $0.39\pm0.03$ & -0.39 & -0.40 \\
GJ4 A & 00:05:41 & 45:48:43 & $88.4\pm 1.6$ & H & K6 &  9.0 & $5.262\pm0.016$ & $0.66\pm0.03$ & 0.10 & -0.05 \\
GJ4 B & 00:05:41 & 45:48:43 & $88.4\pm 1.6$ & H & K7 &  9.0 & $5.284\pm0.023$ & $0.65\pm0.04$ & 0.11 & -0.04 \\
GJ14 & 00:17:06 & 40:56:53 & $66.7\pm 0.9$ & H & M0.5 &  9.0 & $5.577\pm0.024$ & $0.72\pm0.03$ & 0.08 & -0.10 \\
GJ15 A & 00:18:22 & 44:01:22 & $278.8\pm 0.8$ & H & M1 &  8.1 & $4.018\pm0.020$ & $0.41\pm0.03$ & -0.32 & -0.36 \\
GJ15 B & 00:18:25 & 44:01:38 & $278.8\pm 0.8$ & H & M3.5 & 11.1 & $5.948\pm0.024$ & $0.16\pm0.01$ & -0.50 & -0.52 \\
GJ1009 & 00:21:56 & -31:24:21 & $55.6\pm 2.3$ & H & M1.5 & 11.2 & $6.785\pm0.017$ & $0.55\pm0.03$ & 0.35 & 0.11 \\
GJ26 & 00:38:59 & 30:36:58 & $80.1\pm 3.9$ & Y & M2.5 & 11.1 & $6.606\pm0.029$ & $0.43\pm0.03$ & 0.09 & -0.08 \\
GJ27.1 & 00:39:58 & -44:15:11 & $41.7\pm 2.8$ & H & M0.5 & 11.4 & $7.394\pm0.029$ & $0.55\pm0.03$ & 0.06 & -0.09 \\
GJ34 B & 00:49:06 & 57:48:54 & $134.1\pm 0.5$ & H & M0 &  7.5 & $3.881\pm0.490$ & $0.76\pm0.11$ & 0.32 & 0.09 \\
GJ48 & 01:02:32 & 71:40:47 & $121.4\pm 1.2$ & H & M3 & 10.0 & $5.449\pm0.017$ & $0.48\pm0.03$ & 0.28 & 0.05 \\
GJ49 & 01:02:38 & 62:20:42 & $100.4\pm 1.5$ & H & M1.5 &  9.6 & $5.371\pm0.020$ & $0.58\pm0.03$ & 0.26 & 0.06 \\
GJ54.1 & 01:12:30 & -16:59:56 & $268.8\pm 3.2$ & Y & M4.5 & 12.1 & $6.420\pm0.017$ & $0.13\pm0.01$ & -0.33 & -0.43 \\
GJ70 & 01:43:20 & 04:19:18 & $87.6\pm 2.0$ & H & M2 & 10.9 & $6.516\pm0.023$ & $0.41\pm0.03$ & -0.02 & -0.15 \\
GJ83.1 & 02:00:12 & 13:03:11 & $224.8\pm 2.9$ & Y & M4.5 & 12.3 & $6.648\pm0.017$ & $0.14\pm0.01$ & -0.25 & -0.35 \\
GJ3126 & 02:01:35 & 63:46:12 & $78.4\pm10.6$ & Y & M3 & 11.0 & $6.389\pm0.018$ & $0.48\pm0.03$ & 0.39 & 0.12 \\
GJ87 & 02:12:20 & 03:34:32 & $96.0\pm 1.7$ & H & M1.5 & 10.0 & $6.077\pm0.020$ & $0.45\pm0.03$ & -0.26 & -0.32 \\
GJ96 & 02:22:14 & 47:52:48 & $83.8\pm 1.1$ & H & M0.5 &  9.4 & $5.554\pm0.026$ & $0.62\pm0.03$ & 0.11 & -0.05 \\
GJ105 B & 02:36:15 & 06:52:18 & $139.3\pm 0.5$ & H & M3.5 & 11.7 & $6.574\pm0.020$ & $0.25\pm0.02$ & 0.00 & -0.13 \\
GJ109 & 02:44:15 & 25:31:24 & $133.2\pm 2.3$ & H & M3 & 10.6 & $5.961\pm0.021$ & $0.35\pm0.03$ & -0.06 & -0.18 \\
GJ156 & 03:54:35 & -06:49:33 & $64.2\pm 1.1$ & H & M0 &  9.0 & $5.629\pm0.024$ & $0.73\pm0.03$ & 0.08 & -0.10 \\
GJ169 & 04:29:00 & 21:55:21 & $87.8\pm 1.0$ & H & K7 &  8.3 & $4.875\pm0.016$ & $0.74\pm0.03$ & 0.14 & -0.05 \\
GJ172 & 04:37:40 & 52:53:37 & $98.9\pm 1.0$ & H & K8 &  8.6 & $5.047\pm0.018$ & $0.65\pm0.04$ & -0.00 & -0.14 \\
GJ173 & 04:37:41 & -11:02:19 & $90.1\pm 1.7$ & H & M1.5 & 10.3 & $6.091\pm0.021$ & $0.48\pm0.03$ & 0.04 & -0.11 \\
GJ176 & 04:42:55 & 18:57:29 & $107.8\pm 2.9$ & H & M2 &  9.9 & $5.607\pm0.034$ & $0.50\pm0.03$ & 0.17 & -0.02 \\
GJ179 & 04:52:05 & 06:28:35 & $81.4\pm 4.0$ & H & M3.5 & 11.9 & $6.942\pm0.018$ & $0.36\pm0.02$ & 0.34 & 0.08 \\
GJ180 & 04:53:49 & -17:46:24 & $82.5\pm 2.4$ & H & M2 & 10.9 & $6.598\pm0.021$ & $0.42\pm0.03$ & -0.09 & -0.20 \\
GJ3325 & 05:03:20 & -17:22:24 & $108.6\pm 2.7$ & H & M3 & 11.7 & $6.936\pm0.021$ & $0.27\pm0.02$ & -0.22 & -0.28 \\
GJ191 & 05:11:40 & -45:01:06 & $255.7\pm 0.9$ & H & M1.0 &  8.8 & $5.049\pm0.021$ & $0.27\pm0.03$ & -1.01 & -0.82 \\
GJ192 & 05:12:42 & 19:39:56 & $81.3\pm 4.1$ & H & M2 & 10.8 & $6.470\pm0.024$ & $0.45\pm0.03$ & 0.04 & -0.11 \\
GJ205 & 05:31:27 & -03:40:38 & $176.8\pm 1.2$ & H & M1.5 &  8.0 & $3.870\pm0.030$ & $0.63\pm0.03$ & 0.32 & 0.11 \\
GJ3356 & 05:34:52 & 13:52:46 & $80.6\pm 9.8$ & Y & M3.5 & 11.8 & $6.936\pm0.016$ & $0.37\pm0.02$ & 0.25 & 0.02 \\
GJ208 & 05:36:30 & 11:19:40 & $89.0\pm 1.0$ & H & M0 &  8.8 & $5.269\pm0.023$ & $0.65\pm0.04$ & -0.04 & -0.17 \\
GJ212 & 05:41:30 & 53:29:23 & $80.4\pm 1.7$ & H & M0.5 &  9.8 & $5.759\pm0.016$ & $0.60\pm0.03$ & 0.18 & 0.00 \\
GJ213 & 05:42:09 & 12:29:21 & $171.7\pm 1.1$ & G08 & M4 & 11.6 & $6.389\pm0.016$ & $0.22\pm0.02$ & -0.11 & -0.21 \\
GJ3378 & 06:01:11 & 59:35:49 & $132.1\pm 4.9$ & Y & M3.5 & 11.7 & $6.639\pm0.018$ & $0.25\pm0.02$ & -0.02 & -0.14 \\
GJ & 06:07:43 & -25:44:41 & $88.1\pm 2.5$ & H & n/a & 11.9 & $7.169\pm0.023$ & $0.30\pm0.02$ & -0.14 & -0.23 \\
GJ226 & 06:10:19 & 82:06:24 & $106.7\pm 1.3$ & H & M2 & 10.5 & $6.061\pm0.018$ & $0.41\pm0.03$ & -0.00 & -0.14 \\
GJ229 & 06:10:34 & -21:51:52 & $173.8\pm 1.0$ & H & M0.5 &  8.1 & $4.150\pm0.030$ & $0.58\pm0.03$ & 0.11 & -0.05 \\
GJ239 & 06:37:10 & 17:33:53 & $102.6\pm 1.6$ & H & M0 &  9.6 & $5.862\pm0.024$ & $0.47\pm0.03$ & -0.40 & -0.43 \\
GJ250 B & 06:52:18 & -05:11:25 & $114.8\pm 0.4$ & H & M2 & 10.1 & $5.723\pm0.036$ & $0.45\pm0.03$ & 0.05 & -0.10 \\
GJ251 & 06:54:48 & 33:16:05 & $179.0\pm 1.6$ & H & M3 &  9.9 & $5.275\pm0.023$ & $0.35\pm0.03$ & -0.02 & -0.15 \\
GJ273 & 07:27:24 & 05:13:32 & $267.4\pm 0.8$ & G08 & M3.5 &  9.9 & $4.857\pm0.023$ & $0.29\pm0.02$ & 0.08 & -0.09 \\
GJ1097 & 07:28:45 & -03:17:53 & $81.4\pm 2.5$ & H & M3 & 11.5 & $6.704\pm0.027$ & $0.40\pm0.03$ & 0.27 & 0.04 \\
GJ277.1 & 07:34:27 & 62:56:29 & $87.2\pm 2.3$ & H & M0.5 & 10.4 & $6.556\pm0.018$ & $0.40\pm0.03$ & -0.50 & -0.49 \\
GJ3459 & 07:38:40 & -21:13:28 & $94.3\pm 3.3$ & H & M3 & 11.7 & $7.063\pm0.023$ & $0.29\pm0.02$ & -0.24 & -0.29 \\
GJ285 & 07:44:40 & 03:33:08 & $167.9\pm 2.3$ & H & M4.5 & 11.2 & $5.698\pm0.017$ & $0.31\pm0.02$ & 0.58 & 0.27 \\
GJ2066 & 08:16:07 & 01:18:09 & $109.6\pm 1.5$ & H & M2 & 10.1 & $5.766\pm0.024$ & $0.46\pm0.03$ & 0.05 & -0.11 \\
GJ308.1 & 08:29:56 & 61:43:32 & $50.7\pm 1.8$ & H & M0 & 10.3 & $6.781\pm0.017$ & $0.59\pm0.03$ & -0.20 & -0.30 \\
GJ310 & 08:36:25 & 67:17:42 & $72.6\pm 1.3$ & H & M1 &  9.3 & $5.580\pm0.015$ & $0.68\pm0.03$ & 0.16 & -0.01 \\
GJ317 & 08:40:59 & -23:27:22 & $65.3\pm 0.4$ & AE12 & M3.5 & 12.0 & $7.028\pm0.020$ & $0.43\pm0.03$ & 0.50 & 0.19 \\
GJ324 B & 08:52:40 & 28:18:59 & $81.0\pm 0.8$ & H & M4 & 13.2 & $7.666\pm0.023$ & $0.26\pm0.02$ & 0.34 & 0.11 \\
GJ338 A & 09:14:22 & 52:41:11 & $162.8\pm 2.9$ & Y & M0 &  7.6 & $3.988\pm0.036$ & $0.65\pm0.04$ & 0.04 & -0.10 \\
GJ338 B & 09:14:24 & 52:41:11 & $162.8\pm 2.9$ & Y & M0 &  7.7 & $4.136\pm0.020$ & $0.62\pm0.04$ & -0.11 & -0.22 \\
GJ1125 & 09:30:44 & +00:19:21 & $103.5\pm 3.9$ & H & M3.5 & 11.7 & $6.871\pm0.024$ & $0.29\pm0.02$ & -0.07 & -0.18 \\
GJ353 & 09:31:56 & 36:19:12 & $71.9\pm 1.8$ & H & M0 & 10.2 & $6.302\pm0.020$ & $0.53\pm0.03$ & -0.10 & -0.20 \\
GJ357 & 09:36:01 & -21:39:38 & $110.8\pm 1.9$ & H & M2.5 & 10.9 & $6.475\pm0.017$ & $0.33\pm0.03$ & -0.26 & -0.31 \\
GJ361 & 09:41:10 & 13:12:34 & $88.8\pm 1.7$ & H & M1.5 & 10.4 & $6.128\pm0.020$ & $0.48\pm0.03$ & 0.04 & -0.11 \\
GJ362 & 09:42:51 & 70:02:21 & $88.1\pm 2.4$ & H & M3 & 11.2 & $6.469\pm0.016$ & $0.42\pm0.03$ & 0.27 & 0.03 \\
GJ373 & 09:56:08 & 62:47:18 & $94.7\pm 1.3$ & H & M0 &  9.0 & $5.200\pm0.024$ & $0.64\pm0.04$ & 0.11 & -0.04 \\
GJ380 & 10:11:22 & 49:27:15 & $205.2\pm 0.5$ & H & K7 &  6.6 & $3.210\pm0.030$ & $0.71\pm0.03$ & 0.02 & -0.14 \\
GJ382 & 10:12:17 & -03:44:44 & $127.1\pm 1.9$ & H & M1.5 &  9.3 & $5.015\pm0.020$ & $0.54\pm0.03$ & 0.22 & 0.02 \\
GJ388 & 10:19:36 & 19:52:12 & $204.6\pm 2.8$ & Y & M3 &  9.4 & $4.593\pm0.017$ & $0.42\pm0.03$ & 0.37 & 0.10 \\
GJ390 & 10:25:10 & -10:13:43 & $81.0\pm 1.9$ & H & M1 & 10.2 & $6.032\pm0.017$ & $0.54\pm0.03$ & 0.09 & -0.06 \\
GJ393 & 10:28:55 & +00:50:27 & $141.5\pm 2.2$ & H & M2 &  9.7 & $5.311\pm0.023$ & $0.44\pm0.03$ & 0.01 & -0.14 \\
GJ394 & 10:30:25 & 55:59:56 & $74.9\pm 5.6$ & Y & K7 &  8.7 & $5.361\pm0.016$ & $0.71\pm0.03$ & 0.01 & -0.16 \\
GJ397 & 10:31:24 & 45:31:33 & $63.5\pm 1.1$ & H & K7 &  8.8 & $5.564\pm0.024$ & $0.75\pm0.03$ & 0.07 & -0.13 \\
GJ402 & 10:50:52 & 06:48:29 & $147.9\pm 3.5$ & H & M4 & 11.6 & $6.371\pm0.016$ & $0.26\pm0.02$ & 0.16 & -0.02 \\
GJ406 & 10:56:28 & 07:00:53 & $419.1\pm 2.1$ & Y & M5.5 & 13.5 & $6.084\pm0.017$ & $0.10\pm0.00$ & 0.43 & 0.19 \\
GJ408 & 11:00:04 & 22:49:58 & $150.1\pm 1.7$ & H & M2.5 & 10.0 & $5.540\pm0.030$ & $0.37\pm0.03$ & -0.07 & -0.19 \\
GJ410 & 11:02:38 & 21:58:01 & $85.0\pm 1.1$ & H & M0 &  9.6 & $5.688\pm0.021$ & $0.59\pm0.03$ & 0.04 & -0.10 \\
GJ411 & 11:03:20 & 35:58:11 & $392.6\pm 0.7$ & H & M2 &  7.5 & $3.360\pm0.030$ & $0.39\pm0.03$ & -0.32 & -0.35 \\
GJ412 A & 11:05:28 & 43:31:36 & $206.3\pm 1.0$ & H & M0.5 &  8.8 & $4.769\pm0.020$ & $0.39\pm0.03$ & -0.39 & -0.40 \\
GJ413.1 & 11:09:31 & -24:35:55 & $93.0\pm 1.7$ & H & M2 & 10.4 & $6.097\pm0.023$ & $0.46\pm0.03$ & 0.08 & -0.08 \\
GJ414 A & 11:11:05 & 30:26:45 & $84.2\pm 0.9$ & H & K9 &  8.3 & $4.979\pm0.018$ & $0.74\pm0.03$ & 0.08 & -0.11 \\
GJ414 B & 11:11:02 & 30:26:41 & $84.2\pm 0.9$ & H & M1.5 & 10.0 & $5.734\pm0.020$ & $0.58\pm0.03$ & 0.32 & 0.10 \\
GJ424 & 11:20:04 & 65:50:47 & $112.1\pm 1.0$ & H & M0 &  9.3 & $5.534\pm0.017$ & $0.49\pm0.03$ & -0.29 & -0.35 \\
GJ433 & 11:35:26 & -32:32:23 & $112.6\pm 1.4$ & H & M1.5 &  9.8 & $5.623\pm0.021$ & $0.47\pm0.03$ & -0.02 & -0.15 \\
GJ1148 & 11:41:44 & 42:45:07 & $90.1\pm 2.8$ & H & M4 & 11.9 & $6.822\pm0.016$ & $0.35\pm0.02$ & 0.32 & 0.07 \\
GJ436 & 11:42:11 & 26:42:23 & $98.6\pm 2.3$ & H & M2.5 & 10.7 & $6.073\pm0.016$ & $0.44\pm0.03$ & 0.24 & 0.02 \\
GJ445 & 11:47:41 & 78:41:28 & $186.9\pm 1.7$ & H & M3.5 & 10.8 & $5.954\pm0.027$ & $0.25\pm0.02$ & -0.25 & -0.30 \\
GJ447 & 11:47:44 & +00:48:16 & $298.2\pm 1.7$ & Y & M4 & 11.1 & $5.654\pm0.024$ & $0.17\pm0.01$ & -0.14 & -0.24 \\
GJ450 & 11:51:07 & 35:16:19 & $116.5\pm 1.2$ & H & M1 &  9.8 & $5.606\pm0.017$ & $0.46\pm0.03$ & -0.08 & -0.19 \\
GJ3708  & 12:11:11 & -19:57:38 & $79.4\pm 2.4$ & H & M3 & 11.7 & $7.044\pm0.016$ & $0.35\pm0.03$ & -0.01 & -0.15 \\
GJ3709  & 12:11:16 & -19:58:21 & $79.4\pm 2.4$ & H & M3.5 & 12.6 & $7.777\pm0.000$ & $0.25\pm0.02$ & -0.23 & -0.29 \\
GJ465 & 12:24:52 & -18:14:32 & $113.0\pm 2.5$ & H & M2 & 11.3 & $6.950\pm0.021$ & $0.26\pm0.02$ & -0.65 & -0.56 \\
GJ486 & 12:47:56 & 09:45:05 & $119.5\pm 2.7$ & H & M3.5 & 11.4 & $6.362\pm0.018$ & $0.32\pm0.02$ & 0.23 & 0.01 \\
GJ488 & 12:50:43 & -00:46:05 & $94.6\pm 0.8$ & H & M0.5 &  8.5 & $4.882\pm0.020$ & $0.71\pm0.03$ & 0.17 & -0.01 \\
GJ494 & 13:00:46 & 12:22:32 & $85.5\pm 1.5$ & H & M0.5 &  9.8 & $5.578\pm0.016$ & $0.61\pm0.03$ & 0.34 & 0.12 \\
GJ514 & 13:29:59 & 10:22:37 & $130.6\pm 1.1$ & H & M0.5 &  9.0 & $5.036\pm0.027$ & $0.53\pm0.03$ & -0.03 & -0.15 \\
GJ519 & 13:37:28 & 35:43:03 & $91.4\pm 1.2$ & H & M0 &  9.1 & $5.486\pm0.021$ & $0.60\pm0.03$ & -0.15 & -0.25 \\
GJ526 & 13:45:43 & 14:53:29 & $185.5\pm 1.1$ & H & M1.5 &  8.5 & $4.415\pm0.017$ & $0.50\pm0.03$ & -0.07 & -0.18 \\
GJ3804 & 13:45:50 & -17:58:05 & $97.6\pm 5.0$ & H & M3.5 & 11.9 & $6.902\pm0.044$ & $0.31\pm0.02$ & 0.12 & -0.06 \\
GJ536 & 14:01:03 & -02:39:17 & $99.7\pm 1.6$ & H & M1 &  9.7 & $5.683\pm0.020$ & $0.52\pm0.03$ & -0.04 & -0.16 \\
GJ552 & 14:29:29 & 15:31:57 & $71.4\pm 2.1$ & H & M2 & 10.7 & $6.393\pm0.018$ & $0.52\pm0.03$ & 0.18 & -0.01 \\
GJ553.1 & 14:31:01 & -12:17:45 & $92.4\pm 3.9$ & H & M3.5 & 11.9 & $6.961\pm0.021$ & $0.32\pm0.02$ & 0.14 & -0.05 \\
GJ555 & 14:34:16 & -12:31:10 & $158.5\pm 2.6$ & J05 & M3.5 & 11.3 & $5.939\pm0.034$ & $0.29\pm0.02$ & 0.40 & 0.14 \\
GJ9492 & 14:42:21 & 66:03:20 & $93.2\pm 1.3$ & H & M1.5 & 10.9 & $6.491\pm0.024$ & $0.39\pm0.03$ & -0.10 & -0.21 \\
GJ569 A & 14:54:29 & 16:06:03 & $103.6\pm 1.7$ & H & M2.5 & 10.2 & $5.770\pm0.018$ & $0.48\pm0.03$ & 0.16 & -0.03 \\
GJ570 B & 14:57:26 & -21:24:41 & $169.7\pm 1.0$ & S99 & M1 &  8.0 & $4.246\pm0.033$ & $0.57\pm0.03$ & -0.08 & -0.19 \\
GJ581 & 15:19:26 & -07:43:20 & $160.9\pm 2.6$ & H & M3 & 10.6 & $5.837\pm0.023$ & $0.30\pm0.02$ & -0.10 & -0.20 \\
GJ617 A & 16:16:42 & 67:14:19 & $93.6\pm 0.9$ & H & M1 &  8.6 & $4.953\pm0.018$ & $0.70\pm0.03$ & 0.17 & -0.00 \\
GJ617 B & 16:16:45 & 67:15:22 & $93.1\pm 1.5$ & H & M3 & 10.7 & $6.066\pm0.020$ & $0.47\pm0.03$ & 0.34 & 0.09 \\
GJ623 A & 16:24:09 & 48:21:10 & $124.1\pm 1.2$ & H & M2.5 & 10.3 & $5.915\pm0.023$ & $0.38\pm0.03$ & -0.15 & -0.24 \\
GJ625 & 16:25:24 & 54:18:14 & $153.5\pm 1.0$ & H & M1.5 & 10.1 & $5.833\pm0.024$ & $0.32\pm0.03$ & -0.42 & -0.41 \\
GJ628 & 16:30:18 & -12:39:45 & $233.0\pm 1.6$ & H & M3.5 & 10.1 & $5.075\pm0.024$ & $0.30\pm0.02$ & 0.11 & -0.06 \\
GJ638 & 16:45:06 & 33:30:33 & $102.0\pm 0.7$ & H & K7 &  8.1 & $4.712\pm0.021$ & $0.71\pm0.03$ & 0.03 & -0.13 \\
GJ649 & 16:58:08 & 25:44:39 & $96.7\pm 1.4$ & H & M1 &  9.7 & $5.624\pm0.016$ & $0.54\pm0.03$ & 0.07 & -0.08 \\
GJ655 & 17:07:07 & 21:33:14 & $74.8\pm 3.1$ & H & M3 & 11.6 & $7.042\pm0.016$ & $0.38\pm0.03$ & 0.01 & -0.13 \\
GJ3992 & 17:11:34 & 38:26:33 & $83.3\pm 2.0$ & H & M3.5 & 11.5 & $6.801\pm0.021$ & $0.38\pm0.03$ & 0.17 & -0.03 \\
GJ667 C & 17:18:58 & -34:59:48 & $138.0\pm 0.6$ & F00 & M1.5 & 10.2 & $6.036\pm0.020$ & $0.32\pm0.03$ & -0.50 & -0.47 \\
GJ671 & 17:19:52 & 41:42:49 & $80.8\pm 1.7$ & H & M2.5 & 11.4 & $6.915\pm0.018$ & $0.37\pm0.03$ & -0.11 & -0.21 \\
GJ673 & 17:25:45 & 02:06:41 & $129.9\pm 0.7$ & H & K7 &  7.5 & $4.170\pm0.030$ & $0.71\pm0.03$ & 0.03 & -0.14 \\
GJ678.1 & 17:30:22 & 05:32:54 & $100.2\pm 1.1$ & H & M0 &  9.3 & $5.422\pm0.029$ & $0.57\pm0.03$ & 0.01 & -0.12 \\
GJ687 & 17:36:25 & 68:20:20 & $220.8\pm 0.9$ & H & M3 &  9.2 & $4.548\pm0.021$ & $0.40\pm0.03$ & 0.12 & -0.06 \\
GJ686 & 17:37:53 & 18:35:30 & $123.7\pm 1.6$ & H & M1 &  9.6 & $5.572\pm0.020$ & $0.44\pm0.03$ & -0.25 & -0.31 \\
GJ694 & 17:43:55 & 43:22:43 & $105.5\pm 1.2$ & H & M2.5 & 10.5 & $5.964\pm0.020$ & $0.44\pm0.03$ & 0.16 & -0.03 \\
GJ2130  & 17:46:12 & -32:06:12 & $71.5\pm 2.6$ & H06 & M1.5 & 10.5 & $6.251\pm0.026$ & $0.55\pm0.03$ & 0.23 & 0.03 \\
GJ699 & 17:57:48 & 04:41:36 & $545.4\pm 0.3$ & B99 & M4 &  9.6 & $4.524\pm0.020$ & $0.16\pm0.01$ & -0.59 & -0.58 \\
GJ701 & 18:05:07 & -03:01:52 & $128.9\pm 1.4$ & H & M1 &  9.4 & $5.306\pm0.021$ & $0.48\pm0.03$ & -0.12 & -0.22 \\
GJ4048  & 18:18:04 & 38:46:34 & $88.4\pm 3.6$ & Y & M3 & 11.9 & $7.222\pm0.020$ & $0.29\pm0.02$ & -0.23 & -0.29 \\
GJ4070 & 18:41:59 & 31:49:49 & $87.4\pm 2.7$ & H & M3 & 11.3 & $6.722\pm0.020$ & $0.37\pm0.03$ & -0.01 & -0.15 \\
GJ725 A & 18:42:46 & 59:37:49 & $280.2\pm 2.2$ & H & M3 &  8.9 & $4.432\pm0.020$ & $0.33\pm0.03$ & -0.22 & -0.28 \\
GJ725 B & 18:42:46 & 59:37:36 & $289.5\pm 3.2$ & H & M3.5 &  9.7 & $5.000\pm0.023$ & $0.25\pm0.02$ & -0.38 & -0.39 \\
GJ729 & 18:49:49 & -23:50:10 & $336.7\pm 2.0$ & H & M3.5 & 10.5 & $5.370\pm0.016$ & $0.17\pm0.01$ & -0.41 & -0.44 \\
GJ745 A & 19:07:05 & 20:53:17 & $117.5\pm 2.3$ & H & M1.5 & 10.8 & $6.521\pm0.021$ & $0.30\pm0.03$ & -0.52 & -0.48 \\
GJ745 B & 19:07:13 & 20:52:37 & $114.2\pm 2.3$ & H & M2 & 10.8 & $6.517\pm0.023$ & $0.31\pm0.03$ & -0.49 & -0.46 \\
GJ752 A & 19:16:55 & 05:10:08 & $170.4\pm 1.0$ & H & M2.5 &  9.1 & $4.673\pm0.020$ & $0.48\pm0.03$ & 0.23 & 0.02 \\
GJ1245  & 19:53:54 & 44:24:54 & $220.2\pm 1.0$ & Y & M5.5 & 14.0 & $7.387\pm0.018$ & $0.11\pm0.00$ & -0.07 & -0.18 \\
GJ786 & 20:10:52 & 77:14:20 & $59.1\pm 0.7$ & H & K7 &  8.9 & $5.667\pm0.016$ & $0.76\pm0.03$ & 0.06 & -0.15 \\
GJ793 & 20:30:32 & 65:26:58 & $125.1\pm 1.1$ & H & M2.5 & 10.6 & $5.933\pm0.023$ & $0.38\pm0.03$ & 0.06 & -0.10 \\
GJ806 & 20:45:04 & 44:29:56 & $81.2\pm 1.7$ & H & M1.5 & 10.8 & $6.533\pm0.016$ & $0.44\pm0.03$ & -0.07 & -0.19 \\
GJ & 20:52:33 & -16:58:29 & $175.0\pm 3.4$ & H & M4 & 11.5 & $6.199\pm0.021$ & $0.23\pm0.02$ & 0.04 & -0.10 \\
GJ809 & 20:53:19 & 62:09:15 & $141.9\pm 0.6$ & H & M0.5 &  8.6 & $4.618\pm0.024$ & $0.58\pm0.03$ & 0.06 & -0.09 \\
GJ820 B & 21:06:55 & 38:44:31 & $285.9\pm 0.5$ & H & K7 &  6.0 & $2.700\pm0.030$ & $0.66\pm0.04$ & -0.12 & -0.25 \\
GJ821 & 21:09:17 & -13:18:09 & $82.2\pm 2.2$ & H & M1 & 10.9 & $6.909\pm0.029$ & $0.36\pm0.03$ & -0.54 & -0.51 \\
GJ846 & 22:02:10 & 01:24:00 & $97.6\pm 1.5$ & H & M0 &  9.2 & $5.322\pm0.023$ & $0.60\pm0.03$ & 0.05 & -0.09 \\
GJ849 & 22:09:40 & -04:38:26 & $109.9\pm 2.1$ & H & M3.5 & 10.4 & $5.594\pm0.017$ & $0.49\pm0.03$ & 0.54 & 0.22 \\
GJ851 & 22:11:30 & 18:25:34 & $86.1\pm 1.4$ & H & M2 & 10.2 & $5.823\pm0.016$ & $0.55\pm0.03$ & 0.40 & 0.14 \\
GJ860 A & 22:27:59 & 57:41:45 & $249.9\pm 1.9$ & H & M3 &  9.8 & $4.777\pm0.029$ & $0.32\pm0.02$ & 0.25 & 0.03 \\
GJ873 & 22:46:49 & 44:20:02 & $199.0\pm 0.9$ & G98 & M3.5 & 10.2 & $5.299\pm0.024$ & $0.32\pm0.02$ & 0.11 & -0.07 \\
GJ876 & 22:53:16 & -14:15:49 & $214.6\pm 0.2$ & B02 & M4 & 10.2 & $5.010\pm0.021$ & $0.33\pm0.02$ & 0.40 & 0.12 \\
GJ880 & 22:56:34 & 16:33:12 & $146.1\pm 1.0$ & H & M1.5 &  8.7 & $4.523\pm0.016$ & $0.58\pm0.03$ & 0.25 & 0.05 \\
GJ884 & 23:00:16 & -22:31:27 & $121.7\pm 0.7$ & H & K7 &  7.9 & $4.478\pm0.016$ & $0.68\pm0.03$ & -0.05 & -0.19 \\
GJ887 & 23:05:52 & -35:51:11 & $305.3\pm 0.7$ & H & M0.5 &  7.3 & $3.380\pm0.030$ & $0.49\pm0.03$ & -0.15 & -0.24 \\
GJ891 & 23:10:15 & -25:55:52 & $62.2\pm 3.3$ & H & M2 & 11.3 & $6.995\pm0.021$ & $0.46\pm0.03$ & 0.01 & -0.13 \\
GJ4333 & 23:21:37 & 17:17:25 & $91.0\pm 2.9$ & H & M4 & 11.7 & $6.507\pm0.016$ & $0.40\pm0.02$ & 0.61 & 0.26 \\
GJ895 & 23:24:30 & 57:51:15 & $77.2\pm 1.3$ & H & M1 & 10.0 & $5.871\pm0.021$ & $0.59\pm0.03$ & 0.28 & 0.07 \\
GJ905 & 23:41:54 & 44:10:40 & $316.0\pm 1.1$ & Y & M5 & 12.3 & $5.929\pm0.020$ & $0.14\pm0.01$ & 0.17 & 0.05 \\
GJ908 & 23:49:12 & 02:24:04 & $167.3\pm 1.2$ & H & M1 &  9.0 & $5.043\pm0.020$ & $0.42\pm0.03$ & -0.39 & -0.41 \\
GJ911 & 23:54:46 & -21:46:28 & $41.2\pm 2.6$ & H & M0.5 & 10.8 & $7.117\pm0.034$ & $0.62\pm0.04$ & -0.03 & -0.15 \\
\hline
\end{longtable}

\raggedright{
\small
$\pi$ src: (H) revised Hipparcos catalog \citep{van_Leeuwen-2007}; (Y) \citep{van_Altena-1995}; (G08) \citep{Gatewood-2008};
(AE12) \citep{Anglada-Escude-2012};
(J05) \citep{Jao-2005};
(S99) \citep{Soderhjelm-1999};
(F00) \citep{Fabricius-2000};
(H06) \citep{Henry-2006};
(B99) \citep{Benedict-1999};
(G98) \citep{Gatewood-1998};
(B02) \citep{Benedict-2002}.
%(C) CNS4 catalog (Jahreiss, private comm.) \\
%$^{\dag}$ Gl803 is a young ($\sim$20 Myr) M dwarf with a circumstellar disk \citep{Kalas-2004}. The equation to determine its mass may not be adequate for this age. \\}
%(R95) \citep{Reid-1995}; 
}

\twocolumn

\appendix

\label{appendix}

\section{A new M dwarf metallicity and effective temperature calibration based on line and feature measurements of HARPS M dwarf spectra}

Here we briefly explain the method that we developed to estimate the metallicity and effective temperature of M dwarfs. A paper regarding the full details of this calibration is in preparation (Neves et al., in prep.).

%The main goal of the new calibration is to investigate the metallicity difference between stars with and without planets. 

The method is based on the measurement of `peak-to-peak' equivalent widths (EW) of lines and features from the spectra of our volume-limited M dwarf HARPS sample and uses existing photometric calibrations for metallicity \citep{Neves-2012}  and effective temperature \citep{Casagrande-2008}, as starting values. Our method achieves an increase in precision of the metallicity and effective temperature but the accuracy of the new scale is tied to the accuracy of the photometric calibrations.

%The \citet{Neves-2012} scale is a refinement of the the \citet{Sc}

%This scale is based on the metallicity photometric calibration of \citet{Neves-2012} (being itself a slight refinement of the one of \citet{Schlaufman-2010}) and the effective temperature calibration of \citet{Casagrande-2008}. %Therefore, we're solely improving the precision of the existing photometric calibration with the aid of measurements of lines and features of M dwarf spectra. 

\subsection{Calibration sample}

From the initial 102 M dwarf star spectra of the \citet{Bonfils-2011} sample we initially chose 62 stars with S/N greater than 100. Seven stars (Gl191, Gl285, Gl388, Gl699, Gl729, Gl803, GJ1125) were then discarded \textit{a posteriori}, due to a bad correlation of the line measurements with either the reference metallicity or temperature scales, that can be attributed to high activity/rotation (Gl191, Gl285, Gl388, Gl729, Gl803) or to a bad value of the radial velocity (GJ1125). We ended up with a sample of 55 stars, shown in Table \ref{caltable} in which we based our calibration. Column 1 shows the star designation, column 2 the initial photometric [Fe/H] from \citet{Neves-2012}, column 3 the calibrated [Fe/H] value, column 4 the initial photometric effective temperature, and column 5 the calibrated $T_{eff}$ value.

\begin{table}[]
\centering
\caption{Calibration sample.}%\textcolor{red}{ACTUALIZAR com novos valores!}}
\label{caltable}
\begin{center}
%\resizebox{9cm}{!}{
\begin{tabular}{l r r r r }

\hline
\hline

star & [Fe/H]$_{N12}$ & [Fe/H]$_{NEW}$ & $T_{eff~C08}$ & $T_{eff~NEW}$ \\
Gl465 & -0.56 & -0.66 & 3365 & 3415 \\
Gl438 & -0.51 & -0.39 & 3506 & 3444 \\
Gl667C & -0.51 & -0.53 & 3460 & 3351 \\
Gl54.1 & -0.46 & -0.40 & 2920 & 2970 \\
Gl887 & -0.36 & -0.24 & 3657 & 3472 \\
Gl1 & -0.37 & -0.45 & 3495 & 3566 \\
Gl908 & -0.37 & -0.44 & 3579 & 3496 \\
Gl357 & -0.33 & -0.34 & 3329 & 3351 \\
Gl686 & -0.31 & -0.37 & 3536 & 3453 \\
Gl87 & -0.30 & -0.31 & 3539 & 3557 \\
Gl447 & -0.28 & -0.18 & 2958 & 3034 \\
Gl693 & -0.28 & -0.30 & 3178 & 3233 \\
Gl213 & -0.25 & -0.11 & 3062 & 3088 \\
Gl674 & -0.22 & -0.25 & 3276 & 3258 \\
LP771-95A & -0.09 & -0.34 & 3028 & 3238 \\
Gl832 & -0.18 & -0.19 & 3426 & 3419 \\
Gl701 & -0.19 & -0.27 & 3498 & 3468 \\
Gl536 & -0.16 & -0.12 & 3542 & 3537 \\
HIP31292 & -0.15 & -0.10 & 3156 & 3169 \\
Gl105B & -0.14 & -0.02 & 3057 & 2987 \\
Gl341 & -0.15 & -0.13 & 3606 & 3582 \\
Gl273 & -0.13 & -0.01 & 3119 & 3107 \\
Gl581 & -0.17 & -0.21 & 3186 & 3209	 \\
Gl526 & -0.15 & -0.20 & 3503 & 3560 \\
Gl433 & -0.15 & -0.17 & 3453 & 3461 \\
GJ2066 & -0.11 & -0.18 & 3372 & 3447 \\
Gl678.1A & -0.13 & -0.11 & 3628 & 3589 \\
Gl413.1 & -0.11 & -0.12 & 3388 & 3376 \\
Gl618A & -0.08 & -0.08 & 3231 & 3253 \\
Gl393 & -0.10 & -0.22 & 3346 & 3391 \\
Gl514 & -0.10 & -0.16 & 3515 & 3524 \\
Gl250B & -0.09 & -0.10 & 3352 & 3416 \\
Gl628 & -0.06 & -0.02 & 3091 & 3055 \\
Gl367 & -0.05 & -0.07 & 3379 & 3392 \\
Gl229 & -0.04 & -0.01 & 3532 & 3662 \\
Gl846 & -0.06 & 0.06 & 3628 & 3616 \\
Gl680 & -0.04 & -0.22 & 3355 & 3403 \\
Gl752A & -0.00 & 0.06 & 3328 & 3369 \\
Gl877 & -0.02 & -0.01 & 3257 & 3296 \\
HIP31293 & 0.01 & -0.04 & 3236 & 3277 \\
Gl569A & 0.00 & -0.08 & 3327 & 3204 \\
Gl588 & 0.03 & 0.07 & 3277 & 3325 \\
Gl205 & -0.01 & 0.22 & 3576 & 3736 \\
Gl358 & 0.04 & -0.01 & 3194 & 3097 \\
Gl551 & 0.07 & -0.00 & 2625 & 2659 \\
Gl176 & 0.03 & -0.01 & 3344 & 3346 \\
Gl382 & 0.05 & 0.04 & 3397 & 3338 \\
Gl300 & 0.06 & 0.14 & 2973 & 2829 \\
Gl479 & 0.06 & 0.02 & 3219 & 3137 \\
Gl880 & 0.08 & 0.07 & 3453 & 3600 \\
Gl682 & 0.10 & 0.11 & 2973 & 2906 \\
Gl555 & 0.11 & 0.17 & 2983 & 2864 \\
Gl876 & 0.14 & 0.15 & 3036 & 2948 \\
LTT9759 & 0.16 & 0.21 & 3317 & 3333 \\
Gl849 & 0.23 & 0.24 & 3170 & 3121 \\

\hline

\hline
\hline
\end{tabular}
%}
\end{center}
\end{table}

\subsection{Method}

From our calibration sample we first measured `peak-to-peak' equivalent widths (EWs) of lines and features using the 26 redder orders of median normalized HARPS spectra, in the region between 530 to 690 nm. Here we consider features as blended lines. We define the `peak-to-peak' equivalent widths as

\begin{equation}
W = \sum{\frac{F_{pp}-F_{\lambda}}{F_{pp}}\Delta\lambda},
\label{ew}
\end{equation}
where $F_{pp}$ is the value of the flux between the peaks of the line/feature at each integration step and $F_{\lambda}$ the flux of the line/feature. The measurement of the EWs is illustrated in Fig. \ref{spec}, where the `peak-to-peak' flux corresponds to the red dotted lines, and the black line is the flux of the reference spectra. The EW is thus measured between the red dotted line and the solid black line. We used the very high S/N ($\sim$1430 @ 550nm) spectral orders of the star Gl 205 as a reference from where the line/feature regions are going to be measured for all other stars. We rejected lines/features with EW $<$ 8 $m\AA$ and very steep lines/features.%(\textcolor{red}{elaborate here?})}.

\begin{figure}[h]
\begin{center}
\includegraphics[scale=0.45]{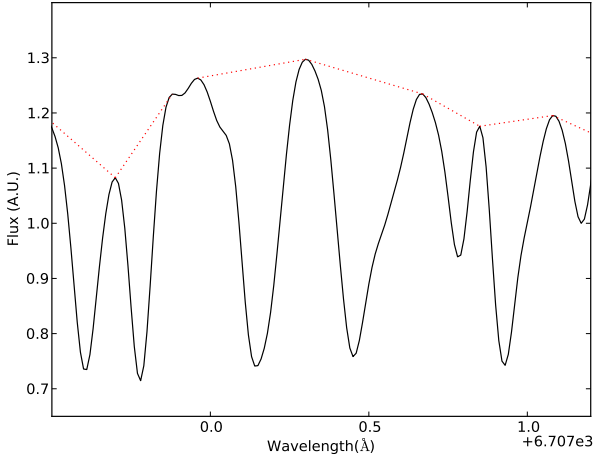}
\end{center}
\caption{Small region of the Gl 205 spectra illustrating the 'peak to peak' equivalent width line measurement. The red dotted line represents the `peak-to-peak' flux.}
\label{spec}
\end{figure}

We investigated the correlations and partial correlations of [Fe/H] and $T_{eff}$ with the line/feature EWs. Fig. \ref{pcorr} shows the histograms of the partial correlation values of [Fe/H] with $T_{eff}$ kept constant(solid blue histogram) and the partial correlation values of $T_{eff}$ with [Fe/H] kept constant (dashed green histogram). We observe that a significant number of lines have a good correlation with the parameters.

\begin{figure}[h]
\begin{center}
\includegraphics[scale=0.45]{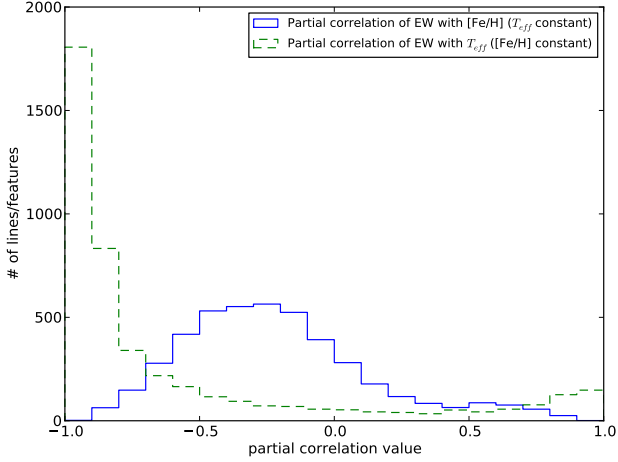}
\end{center}
\caption{Histograms of the partial correlations of [Fe/H] (solid blue histogram) and $T_{eff}$ (dashed green histogram) }
\label{pcorr}
\end{figure}

Then we calculated a linear fit of the EWs with the metallicity (taken from \citet{Neves-2012}) and effective temperature (taken from \citet{Casagrande-2008}), using a least squares approach. For each EW $i$ and for each star $m$ we have,

\begin{equation}
W_{i,m} = \alpha_{i}[Fe/H]_{m}^{T} + \beta_{i}T_{eff m}^{T} + \gamma_{i}, %Ones_{m}^{T},
\label{first}
\end{equation}

where $W_{i,m}$ is a $i\times m$ matrix containing the EWs, and both $[Fe/H]_{m}$, and $T_{eff m}$ are $1\times m$ vectors. The $\alpha$ and the $\beta$ are the coefficients related to metallicity and effective temperature, respectively, while $\gamma$ is an independent coefficient. %The Ones vector is a $1\times m$ vector of unit elements.

The error of each coefficient is calculated as

\begin{equation}
\label{ap:rss}
\epsilon_{i} = \sqrt{RSS.J_{i,i}},
\end{equation}
where RSS is the residual sum of squares, expressed as
\begin{equation}
RSS = \frac{\sum{(x_{i,model}-x_{i})^{2}}}{n_{obs}-n_{coef}},
\end{equation}
and $J_{i,i}$ is the diagonal of the estimate of the jacobian matrix around the solution. The $x_{i,model}$,$x_{i}$, $n_{obs}$, and $n_{coef}$ from Eq. \ref{ap:rss} are, respectively, the predicted value of the data, $x_{i}$, by the regression model, the data values, the number of data points, and the number of coefficients. 

The total error of the coefficients can then be written as 

\begin{equation} 
\epsilon = \sqrt{\epsilon\alpha^{2}+\epsilon\beta^{2}+\epsilon\gamma^{2}}.
\end{equation}
Here we assume that both [Fe/H] and temperature are independent and do not correlate with each other. 

Our aim is to increase the metallicity precision using the photometric calibration as reference. In order to do this, we want to recover the values of the metallicity and temperature by doing a weighted least squares refit. To calculate the weights for the least squares refit we just invert the squared errors of the coefficients, and normalize the expression,

\begin{equation}
E_{i} = \frac{1/\epsilon_{i}^{2}}{\sum{1/\epsilon_{i}^{2}}}.
\label{weight}
\end{equation}

To invert the fit of Eq. \ref{first} we first take the calculated coefficients from the first fit and define the coefficient matrix as

\begin{equation}
C_{i,3} = \left[\begin{array}{ccc} \alpha_{1,1} & \beta_{1,2} & \gamma_{1,3} \\ \alpha_{2,1} & \beta_{2,2} & \gamma_{2,3} \\... & ... & ...\\ \alpha_{i,1} & \beta_{i,2} & \gamma_{i,3} \end{array}\right].
\end{equation}
Then we invert Eq. \ref{first}. After some operations we have 

\begin{equation}
[[Fe/H],Teff,Ind]_{3,m} = (C^{T}_{3,i}C_{i,3})^{-1}C^{T}_{3,i}W_{i,m},
\label{refit}
\end{equation}
where $C^{T}$ is the transpose of $C$ and $Ind$ is the value of the independent parameter. 

Finally, we use a \textit{levenberg-marquardt} algorithm and apply the weights (Eq. \ref{weight}) to Eq. \ref{refit}, recovering one value of metallicity and effective temperature for each star.

We also tried other methods, such as choosing groups of lines with a high correlation or partial correlation coefficients and then applying the same method as described in this Appendix. However, the weighted least squares method using all 4441 lines performed best at minimizing the uncertainties of both metallicity and effective temperature.

%\textcolor{red}{A plot of the best calibration here for Feh and teff?}

Using this method, we get a dispersion of metallicity and effective temperature of 0.08 dex and 80K respectively. Figs. \ref{fehfeh} and \ref{teffteff} show the comparison between the values obtained in this work and the reference calibrations for metallicity and effective temperature, respectively. We emphasize that we only get an improvement of the precision. The accuracy of the calibration, as well as systematic errors, are tied to the original determinations of both [Fe/H] and temperature.

\begin{figure}[h]
\begin{center}
\includegraphics[scale=0.45]{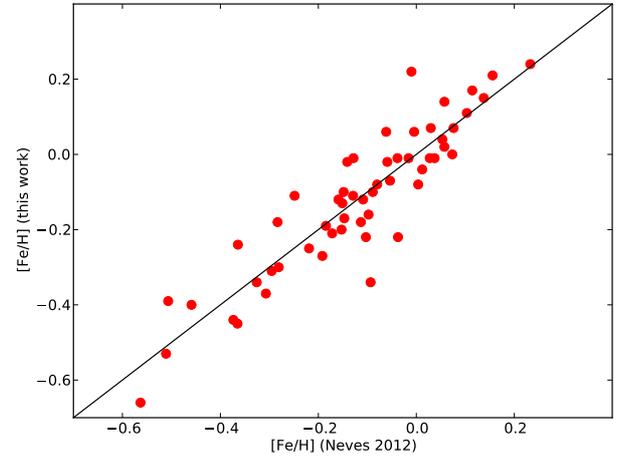}
\end{center}
\caption{[Fe/H] comparison between this work and the photometric calibration of \citet{Neves-2012}.}
\label{fehfeh}
\end{figure}
\begin{figure}[h]
\begin{center}
\includegraphics[scale=0.45]{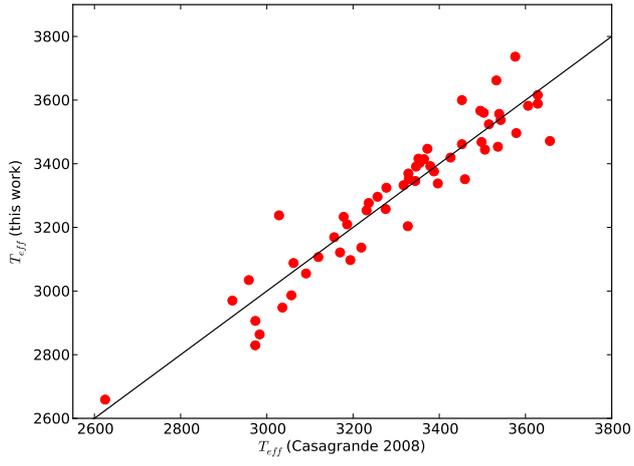}
\end{center}
\caption{$T_{eff}$ comparison between this work and the photometric calibration of \citet{Casagrande-2008}.}
\label{teffteff}
\end{figure}

%%_____________________________________________________________
%%                              Table longer than a single page  
%%  In the preamble, use:              \usepackage{longtable}
%%-------------------------------------------------------------
%%          All long tables have to be placed at the end, after 
%%                                        \end{thebibliography}
%%
%% In the text, at the place where the large table should appear
%% add the command:
%\addtocounter{table}{1}
%% Tables counters will be well numbered.
%%
%\end{thebibliography}
%% If table 2
%\longtab{2}{
%\begin{longtable}{lllrrr}
%\caption{\label{kstars} Sample stars with absolute magnitude}\\
%\hline\hline
%Catalogue& $M_{V}$ & Spectral & Distance & Mode & Count Rate \\
%\hline
%\endfirsthead
%\caption{continued.}\\
%\hline\hline
%Catalogue& $M_{V}$ & Spectral & Distance & Mode & Count Rate \\
%\hline
%\endhead
%\hline
%\endfoot
%%%
%Gl 33    & 6.37 & K2 V & 7.46 & S & 0.043170\\
%Gl 66AB  & 6.26 & K2 V & 8.15 & S & 0.260478\\
%Gl 68    & 5.87 & K1 V & 7.47 & P & 0.026610\\
%         &      &      &      & H & 0.008686\\
%Gl 86 
%\footnote{Source not included in the HRI catalog. See Sect.~5.4.2 for details.}
%         & 5.92 & K0 V & 10.91& S & 0.058230\\
%\end{longtable}
%}

\end{document}